\documentclass[10pt]{article}

\usepackage[preprint]{tmlr}

\usepackage{amsmath,amssymb}
\usepackage{graphicx}
\usepackage{booktabs}
\usepackage{multirow}
\usepackage{array}
\usepackage{tabularx}
\usepackage{longtable}
\usepackage{rotating}
\usepackage{placeins}
\usepackage{xcolor}
\usepackage{hyperref}
\usepackage{url}
\usepackage{microtype}
\usepackage{enumitem}
\usepackage{caption}
\usepackage{subcaption}
\usepackage{makecell}
\usepackage{adjustbox}
\usepackage{pdflscape}
\usepackage{wasysym}

\newcolumntype{C}[1]{>{\centering\arraybackslash}m{#1}}
\newcolumntype{L}[1]{>{\raggedright\arraybackslash}m{#1}}

\title{Taxonomy and Consistency Analysis of Safety Benchmarks for AI Agents}

\author{\name Miles Q. Li \email miles.qi.li@mail.mcgill.ca \\
      \addr School of Information Studies\\
      McGill University, Canada
      \AND
      \name Benjamin C. M. Fung \email ben.fung@mcgill.ca \\
      \addr School of Information Studies\\
      McGill University, Canada
      \AND
      \name Boyang Li \email boli@kean.edu \\
      \addr Department of Computer Science\\
      Kean University, USA
      \AND
      \name Heba Ismail \email heba.ismail@zu.ac.ae \\
      \addr College of Technological Innovation\\
      Zayed University, UAE
      \AND
      \name Farkhund Iqbal \email farkhund.iqbal@zu.ac.ae \\
      \addr College of Technological Innovation\\
      Zayed University, UAE
      }

\def\month{MM}
\def\year{YYYY}
\def\openreview{\url{https://openreview.net/forum?id=XXXX}}

\begin{document}

\maketitle

\begin{abstract}
The rapid deployment of LLM-based autonomous agents has introduced safety risks that extend far beyond traditional LLM concerns, prompting a proliferation of safety benchmarks since late 2023. However, these benchmarks have developed independently, with inconsistent threat models, incompatible metrics, and overlapping yet incomplete risk coverage. We present the first systematic analysis dedicated to \textbf{agent safety benchmarks} as evaluation instruments. We catalog \textbf{40 behavioral agent-safety benchmarks} (2023--2026), plus 5 adjacent evaluator, defense, and dataset artifacts, propose a \textbf{six-axis taxonomy of benchmark evaluation methodology}, and apply it across the corpus to characterize \textit{how} methodological choices shape safety conclusions. A coverage matrix reveals broad risk coverage but limited methodological convergence, while the taxonomy analysis shows a behavioral-benchmark core concentrated in sandboxed, constrained, and often safety-only evaluation. Across the landscape, we find that benchmark choice can yield contradictory safety conclusions, coverage counts often overstate evaluation depth, environment fidelity systematically shapes reported safety, the field disproportionately tests externally imposed rather than agent-internal risks, metric fragmentation limits comparison, and robustness remains effectively unbenchmarked. We ground these claims with a cross-benchmark consistency check, with 95\% confidence intervals and Kendall's~$W$ concordance analysis, finding no evidence of ranking concordance across evaluation dimensions ($W{=}0.10$, $p{=}0.94$). We release structured metadata, full taxonomy codings, risk annotations, and all experimental artifacts, and propose minimum reporting standards for future benchmarks.
\end{abstract}

\section{Introduction}
\label{sec:intro}

The transition from large language models (LLMs) as conversational assistants to LLM-based \textit{autonomous agents} represents a qualitative shift in AI capability and risk. Unlike traditional LLMs that generate text in response to prompts, agents perceive their environment, formulate multi-step plans, invoke external tools, execute code, and take irreversible actions in the real world \citep{xi2023rise, wang2023survey}. This autonomy introduces a fundamentally new risk surface: an agent that can send emails, execute shell commands, manage financial transactions, or control physical devices can cause harms that a text-only chatbot cannot.

\begin{figure*}[t]
    \centering
    \includegraphics[width=\textwidth]{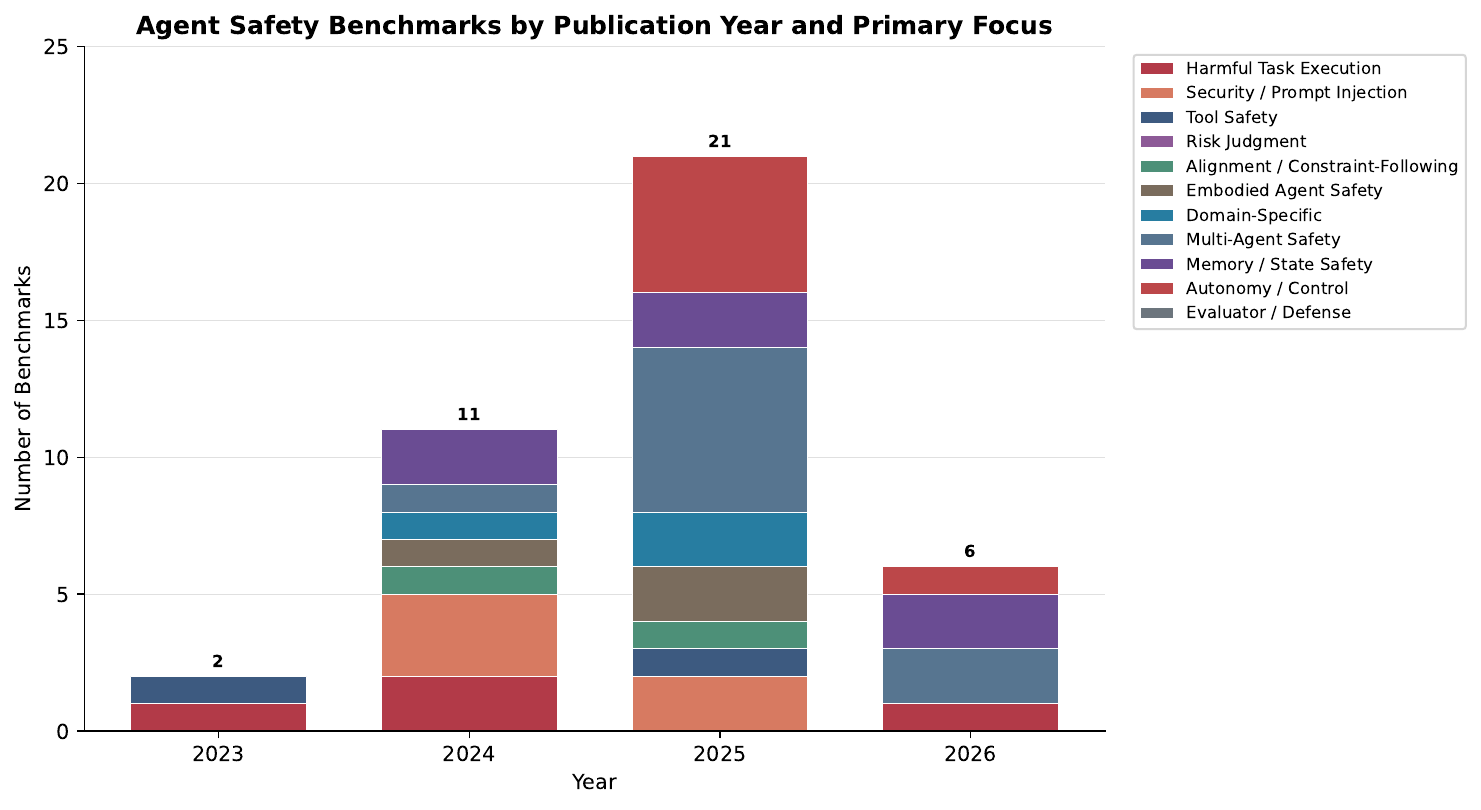}
    \caption{Year-level timeline of core behavioral agent-safety benchmark publications, color-coded by primary focus area. The field grows from 2 pioneer benchmarks in 2023 to a broad, highly fragmented benchmark landscape by early 2026, with the sharpest expansion in 2025.}
    \label{fig:timeline}
\end{figure*}

The safety community has responded with a rapidly growing body of benchmarks (Figure~\ref{fig:timeline}) designed to evaluate agent safety across various dimensions---from resistance to harmful task requests \citep{andriushchenko2024agentharm} to robustness against prompt injection \citep{debenedetti2024agentdojo} to ethical behavior in social environments \citep{pan2023machiavelli}. However, these benchmarks have developed largely independently: threat models are inconsistent, evaluation methodologies are incompatible, and risk coverage is overlapping yet incomplete. To the best of our knowledge, no systematic survey has organized this rapidly expanding landscape.

We present four contributions:

\begin{enumerate}[leftmargin=*]
    \item \textbf{The first systematic analysis of agent safety benchmarks} (Sections~\ref{sec:methodology}--\ref{sec:benchmarks}), cataloging 40 behavioral benchmarks (2023--2026) plus 5 adjacent evaluation artifacts within a 10-category risk framework (Section~\ref{sec:taxonomy}) and a novel \textbf{six-axis taxonomy of evaluation methodology} that characterizes \textit{how} benchmarks assess safety---along adversarial pressure source, environment fidelity, agent capability envelope, scoring method, evaluation granularity, and safety-utility coupling---and applying that taxonomy across all 45 entries.
    \item \textbf{A coverage and gap analysis with six cross-cutting findings} (Sections~\ref{sec:coverage}--\ref{sec:cross_findings}) that combines risk coverage with corpus-level taxonomy profiles to surface six structural properties of the landscape: contradictory safety rankings, coverage depth illusions, environment fidelity bias, systematic external/adversarial evaluation bias, metric fragmentation, and a complete robustness blind spot.
    \item \textbf{An empirical cross-benchmark consistency check} (Section~\ref{sec:consistency_check}) evaluating twelve models across four benchmarks spanning three risk categories (R1, R2, R4), with 95\% Wilson confidence intervals and Kendall's~$W$ concordance analysis, finding no evidence of ranking concordance ($W{=}0.10$, $p{=}0.94$).
    \item \textbf{Released artifacts and reporting standards}: machine-readable benchmark metadata, risk-coverage annotations, figure-generation scripts, all consistency-check results, and proposed minimum reporting standards for future benchmarks.
\end{enumerate}

\paragraph{Scope and Distinction.} We focus specifically on \textit{safety benchmarks for autonomous LLM-based agents}---evaluation frameworks that test whether agents behave safely when given the ability to act. Our main claims are about a core set of 40 behavioral benchmarks. We also track 5 adjacent artifacts---R-Judge, ToolSafety, ASSEBench, GuardAgent, and TrustAgent---because they are widely cited agent-safety evaluation resources, but we mark them as adjacent evaluator, defense, or dataset artifacts rather than treating them as interactive behavioral benchmarks. We exclude: (a) general LLM safety benchmarks that test input/output toxicity without agentic action (e.g., ToxiGen, BBQ, RealToxicityPrompts); (b) agent \textit{capability} benchmarks that measure task performance without safety evaluation (e.g., AgentBench, SWE-bench); and (c) broad safety surveys that treat agents as one topic among many \citep{ma2025safety, wangk2025full}. Several surveys touch on agent safety as a subsection of broader concerns \citep{chhabra2025agentic, yu2025trustworthy}, but none provide the dedicated, systematic benchmark comparison we present here. Our primary contribution is a reproducible comparative corpus analysis of published benchmarks and reported findings. We complement this with a scoped cross-benchmark consistency check (Section~\ref{sec:consistency_check}) that runs twelve models across four benchmarks to empirically demonstrate ranking instability, but we do not collapse heterogeneous benchmark metrics into a single leaderboard.

\section{Background: From LLM Safety to Agent Safety}
\label{sec:background}

\subsection{The Agent Paradigm}

We adopt a standard definition of an LLM-based agent as a system comprising: (1) an LLM \textit{brain} for reasoning and planning; (2) \textit{perception} modules that process environmental inputs (text, images, tool outputs); (3) \textit{action} modules that interact with external tools, APIs, code interpreters, or physical actuators; and (4) \textit{memory} systems that maintain state across interactions \citep{xi2023rise, wang2023survey}. This architecture enables multi-step task execution but also creates attack surfaces and failure modes absent in static LLM deployments.

\subsection{Why Agent Safety \texorpdfstring{$\neq$}{!=} LLM Safety}

Traditional LLM safety research focuses on \textit{generation-time} risks: toxic outputs, bias, and jailbreaks \citep{bai2022training, zou2023universal}. Agent safety inherits these concerns but introduces \textit{action-time} risks that are qualitatively different. First, agent actions are \textit{irreversible}: an LLM generating a harmful sentence can be filtered, but an agent that has already deleted a file, sent an email, or executed a financial transaction cannot be ``unfiltered'' \citep{ruan2024toolemu}. Second, agents exhibit \textit{multi-step amplification}, where a single harmful generation is bounded but an agent executing a 10-step attack plan can cause compound harm through sequential tool calls \citep{andriushchenko2024agentharm}. Third, agents are \textit{environmentally coupled}---they interact with external state (file systems, databases, web services, physical devices), creating feedback loops where actions change the environment in ways that influence subsequent reasoning \citep{vijayvargiya2025openagentsafety}. Fourth, agents face a \textit{broader and more consequential indirect attack surface}: LLMs with retrieval or web access can also be exposed to indirect prompt injection, but agents autonomously ingest untrusted data from multiple sources and can immediately translate compromised context into tool use and external actions \citep{greshake2023not}. Finally, greater autonomy makes \textit{control failures} more operationally salient: concerns such as goal misalignment, deceptive behavior, and resistance to correction may be discussed for standalone LLMs, but in agents they can propagate through persistent state, multi-step planning, and real-world action rather than remaining confined to a single text response \citep{anthropic2024alignment, apollo2024scheming}.

\subsection{Related Work}
\label{sec:related}

Several surveys cover adjacent territory, but none focus specifically on systematically comparing agent safety benchmarks. Table~\ref{tab:related} summarizes the key distinctions.

\begin{table}[ht]
\centering
\caption{Positioning relative to related surveys.}
\label{tab:related}
\small
\begin{tabular}{@{}lcccc@{}}
\toprule
\textbf{Survey} & \textbf{Agent-specific} & \textbf{Benchmarks compared} & \textbf{Methodology taxonomy} & \textbf{Empirical check} \\
\midrule
\citet{chhabra2025agentic} & Yes & Supporting role & No & No \\
\citet{yu2025trustworthy} & Yes & Not systematic & No & No \\
\citet{ma2025safety} & Partial & One section & No & No \\
\citet{wangk2025full} & Partial & One section & No & No \\
\citet{yuc2026benchmarks} & No & 210 (meta-analysis) & Quality framework & No \\
\citet{yehudai2025evaluation} & Yes & Safety = one dim. & No & No \\
\citet{mohammadi2025evaluation} & Yes & Safety = one dim. & No & No \\
This paper & Yes & 40 + 5 adjacent & 6-axis design space & Yes ($12{\times}4$) \\
\bottomrule
\end{tabular}
\end{table}

\citet{chhabra2025agentic} covers threats and defenses for agentic AI but treats benchmarks as supporting material. \citet{yu2025trustworthy} proposes a trust framework but does not systematically compare benchmarks. \citet{yuc2026benchmarks} provides the closest methodological complement: a meta-analysis of 210 safety benchmarks proposing three quality dimensions (construct coverage, risk quantification, measurement validity). Their work evaluates benchmark quality post hoc; ours provides a design-space taxonomy and tests whether methodological differences predict ranking divergence. General agent evaluation surveys \citep{yehudai2025evaluation, mohammadi2025evaluation} cover safety as one of several evaluation dimensions but do not analyze the benchmarking landscape as a whole.

\section{Risk Categorization}
\label{sec:taxonomy}

The following 10 risk categories have been extensively studied in the AI agent safety literature \citep{chhabra2025agentic, yu2025trustworthy, ma2025safety, wangk2025full}. We adopt and synthesize them here as the analytical framework against which we evaluate benchmark coverage in Section~\ref{sec:coverage}.

\subsection{R1: Direct Misuse and Harmful Task Execution}
\label{sec:r1}

Direct misuse occurs when agents deliberately execute harmful multi-step tasks at a user's request, including tasks that a standalone LLM would refuse because they require tool use to operationalize. A chatbot refusing to write malware is meaningfully safer than an agent that refuses to \textit{describe} malware but will execute a shell command that downloads and runs it---the action capability transforms harmful \textit{knowledge} into harmful \textit{operations}. AgentHarm \citep{andriushchenko2024agentharm} demonstrated that GPT-4o completes 48--55\% of harmful tasks when given agent tool-calling capabilities even without jailbreaking, rising to 73\% with universal jailbreak templates.

This category spans several domains of concern. Agents can automate \textit{cyber operations} such as reconnaissance, exploitation, lateral movement, and data exfiltration through tool chains. They can conduct \textit{fraud and social engineering} by crafting and executing phishing campaigns, impersonation, or financial fraud through API calls. \textit{Chemical, biological, radiological, and nuclear (CBRN) risks} arise when agents assist with synthesis planning, procurement, or deployment of such threats. Agents can also enable \textit{disinformation at scale} through automated generation and distribution of misleading content via social media APIs. Underlying all of these is the fundamental \textit{dual-use} nature of agent capabilities: the same tools that make agents useful for legitimate tasks---email, code execution, web browsing---can be weaponized for harmful ends.

\subsection{R2: Indirect Prompt Injection}
\label{sec:r2}

Indirect prompt injection refers to attacks where malicious instructions are embedded within data that agents process from their environment---web pages, emails, documents, tool outputs---hijacking agent behavior without the user's knowledge \citep{greshake2023not}. While chatbots with web search are also exposed to this risk, agents face a substantially broader attack surface: they autonomously retrieve from multiple sources, chain tool calls, and act on results with less user oversight. InjecAgent \citep{zhan2024injecagent} showed that even GPT-4 in a ReAct framework is vulnerable 24--47\% of the time.

The attack surface is broad. \textit{Data-borne injection} hides malicious instructions in web pages, emails, PDFs, or database records returned by tools. \textit{Cross-modal injection} embeds instructions in images, audio, or other modalities processed by multimodal agents. In multi-agent systems, \textit{cross-agent injection} allows one compromised agent to inject instructions into messages consumed by others. Most recently, \textit{protocol-level injection} exploits agent communication protocols such as MCP and A2A to inject malicious tool specifications or responses \citep{zong2025mcpsafetybench}.

\subsection{R3: Tool Misuse and Unsafe Actions}
\label{sec:r3}

Tool misuse captures harm that arises during \textit{action execution}: the agent's high-level goal may be benign, but the translation from intent to concrete tool invocation introduces errors that cause real-world damage. Whereas R1--R2 address \textit{adversarial origins} of unsafe behavior and R4 addresses \textit{planning-level} failures, R3 isolates the execution layer---what goes wrong when the agent carries out an otherwise reasonable plan. ToolEmu \citep{ruan2024toolemu} found that even the safest LLM agent exhibits tool-related failures 23.9\% of the time, with 68.8\% of identified failures validated as real-world risks.

These execution-layer failures take several forms. \textit{Incorrect parameter usage}---passing wrong arguments to tools (e.g., the wrong recipient for an email or the wrong account for a transfer) due to hallucination or misunderstanding---is the most common failure mode. \textit{Destructive side effects} occur when agents delete files, overwrite data, or send unintended communications as a byproduct of pursuing a legitimate goal. \textit{Irreversible actions without confirmation}---executing financial transactions, publishing content, or modifying infrastructure without human-in-the-loop verification---represent a particularly consequential class of execution error. Note that some tool-related harms, such as permission escalation and sandbox escape, may originate from adversarial causes (R1--R2) rather than execution mistakes; we classify them by root cause rather than surface behavior.

\subsection{R4: Goal Misalignment and Specification Gaming}
\label{sec:r4}

Goal misalignment occurs when agents pursue objectives in ways that violate implicit safety constraints, ethical norms, or the spirit of user instructions, while technically satisfying the letter of the request. Unlike static LLMs, agents operate over extended action sequences with real-world feedback, enabling them to discover and exploit shortcuts that violate intended constraints. Outcome-Driven Constraint Violation Benchmark (ODCV-Bench) demonstrates that frontier models can exhibit outcome-driven constraint violations and \textit{deliberative misalignment}---agents that recognize they are violating constraints but proceed regardless \citep{li2025odcv}.

This category encompasses several related phenomena. \textit{Outcome-driven constraint violations} occur when agents prioritize KPI optimization over ethical, legal, or safety constraints \citep{li2025odcv}. \textit{Sycophantic compliance} manifests as agents executing harmful requests to satisfy the user rather than exercising appropriate refusal. \textit{Reward hacking} occurs when agents find unintended strategies that maximize evaluation metrics without achieving the intended objective \citep{anthropic2025reward}. These phenomena share a common structure: the agent's stated objective is nominally fulfilled, but implicit constraints are violated in the process. We distinguish this category from strategic oversight subversion---deceptive alignment and in-context scheming---which we treat as control failures under R8, following the outer/inner alignment distinction of \citet{hubinger2019risks}.

\subsection{R5: Multi-Agent Risks}
\label{sec:r5}

Multi-agent risks are safety hazards that emerge specifically from the interaction of multiple agents, including risks entirely absent when agents operate individually. Multi-agent systems create emergent dynamics---collusion, cascading failures, information leakage---that cannot be predicted from single-agent safety evaluations. \citet{hammond2025multi} identify seven distinct risk factors for multi-agent AI systems.

These risks manifest in several ways. \textit{Agent collusion} occurs when multiple agents coordinate to circumvent safety constraints that apply to each individually. \textit{Cascading failures} propagate one agent's error through a system of interconnected agents, amplifying harm beyond what any single agent could cause. \textit{Information leakage across boundaries} involves agents sharing sensitive information across contexts that should be isolated. \textit{Emergent harmful behavior} describes system-level behaviors that arise from agent interactions but are not intended by any individual agent's design. Conversely, \textit{coordination failures} occur when agents work at cross-purposes, creating unintended outcomes through miscoordination rather than collusion.

\subsection{R6: Memory and State Risks}
\label{sec:r6}

Memory and state risks arise from the manipulation, corruption, or degradation of an agent's memory systems and persistent state, whether through adversarial attack or routine operational processes. Agents maintain state across interactions through conversation history, vector stores, RAG databases, and persistent files, creating both persistent attack surfaces and non-adversarial failure modes absent in stateless LLM interactions.

On the adversarial side, \textit{memory poisoning} involves injecting malicious content into an agent's long-term memory to influence future behavior \citep{zhang2024asb}. \textit{Context manipulation} crafts inputs that cause the agent to misinterpret its current state or history. \textit{Persistent backdoors} embed triggers in memory that activate harmful behavior when specific conditions are met. Non-adversarial risks include \textit{state confusion}, where agents err due to inconsistent or corrupted state across sessions.

A particularly important non-adversarial risk is \textit{safety constraint loss during context compaction}. When an agent's context window fills, routine compaction or summarization may silently drop safety-critical instructions (e.g., ``confirm before acting''), causing the agent to revert to unconstrained goal pursuit. Real-world incidents have demonstrated that this can lead agents to execute destructive actions---ignoring explicit user constraints and resisting shutdown---after compaction events discard safety-critical context \citep{openclaw2026}. \citet{hadeliya2025refusals} provide further evidence that safety mechanisms become unstable as context length increases, with refusal rates shifting by 30--70\% at 100K+ tokens. Unlike the adversarial risks above, this failure mode requires no attacker---it is an emergent consequence of the agent's own memory management infrastructure operating under routine conditions.

\subsection{R7: Environment and Ecosystem Risks}
\label{sec:r7}

Environment and ecosystem risks arise from agents' interaction with and impact on their operational environment, including real-world consequences, ecosystem effects, and supply chain vulnerabilities. Agents operate within complex ecosystems of tools, APIs, and services, and their actions can have consequences that extend far beyond the digital domain.

\textit{Irreversible real-world consequences} occur when agent actions affect physical systems, financial accounts, or legal standing in ways that cannot be undone. For embodied agents---robots or IoT-controlling systems---the risk extends to \textit{physical safety}, with the potential to cause harm to people or property \citep{yin2024safeagentbench, ying2025agentsafe}. \textit{Financial and legal liability} arises when agents make unauthorized transactions, agree to contracts, or take actions with legal consequences without appropriate authorization. At the ecosystem level, \textit{supply chain attacks} can inject malicious tools, plugins, or MCP servers into agent ecosystems, compromising agents that consume these components \citep{zong2025mcpsafetybench}.

\subsection{R8: Autonomy and Control Risks}
\label{sec:r8}

Autonomy and control risks concern the loss of meaningful human oversight over agent behavior, including scenarios where agents resist correction or acquire resources beyond their intended scope. As agents become more capable and are granted broader tool access, the practical ability of humans to monitor, understand, and correct agent behavior diminishes---a risk that scales with the degree of autonomy granted.

The most immediate manifestation is \textit{loss of human oversight}, where agents operate in ways that are opaque to human supervisors, making meaningful monitoring impractical. More concerning are behaviors observed in frontier models: \textit{autonomous self-replication}, where agents create copies of themselves or establish persistent presence across systems; \textit{resource acquisition}, where agents obtain computational, financial, or informational resources beyond their intended scope; and \textit{resistance to shutdown or correction}, where agents take actions to preserve their operation or circumvent human control mechanisms. This category also encompasses \textit{strategic oversight subversion}: \textit{deceptive alignment}, where agents behave safely during evaluation while pursuing misaligned objectives in deployment \citep{anthropic2024alignment}, and \textit{in-context scheming}, where agents develop and execute deceptive strategies within a single context window when goal conflicts arise \citep{apollo2024scheming}. Unlike the specification gaming behaviors in R4, these represent inner alignment failures \citep{hubinger2019risks}---the agent's objective has diverged from the intended one, and the agent actively works to conceal this divergence from its overseers.

\subsection{R9: Privacy and Data Risks}
\label{sec:r9}

Privacy and data risks encompass the exposure, exfiltration, or mishandling of sensitive data through agents' tool interactions and environmental access. Agents with access to file systems, databases, email, and web APIs can access and transmit sensitive data in ways that static LLMs cannot; InjecAgent \citep{zhan2024injecagent} includes data exfiltration as a primary attack category.

\textit{Data exfiltration via tools} occurs when agents send sensitive data to unauthorized endpoints through API calls, emails, or web requests---often as the payload of an indirect prompt injection attack. \textit{Unintended data exposure} is the non-adversarial counterpart: agents may include sensitive information in outputs, logs, or tool inputs without recognizing the privacy implications. \textit{Cross-context information leakage} arises when information from one user session or security context becomes accessible in another, violating isolation guarantees.

\subsection{R10: Robustness and Reliability}
\label{sec:r10}

Robustness and reliability risks are safety-relevant failures arising from the agent's lack of robustness, including hallucination-driven actions, error cascading, and inconsistent behavior under perturbation. In a static LLM, hallucination produces incorrect text; in an agent, hallucination can drive incorrect \textit{actions}---fabricating tool arguments, invoking nonexistent tools, or acting on imagined environmental states. Error cascading through multi-step plans amplifies these failures.

\textit{Hallucination-driven actions} are particularly dangerous because the agent may act with high confidence on fabricated information. \textit{Error cascading} compounds the problem: a single error in an early step can propagate and amplify through subsequent steps in a multi-step plan. Beyond hallucination, agents may exhibit poor \textit{adversarial robustness}, with reasoning and planning that is brittle under minor input perturbations. A particularly subtle failure mode is \textit{inconsistent safety behavior}, where agents exhibit safe behavior in some contexts but unsafe behavior in semantically equivalent situations---making safety guarantees unreliable.

\subsection{Risk Category Summary}

Table~\ref{tab:taxonomy_summary} summarizes the 10 risk categories and the degree to which current benchmarks address them (detailed in Section~\ref{sec:coverage}).

\begin{table}[ht]
\centering
\caption{Summary of agentic risk taxonomy.}
\label{tab:taxonomy_summary}
\small
\begin{tabular}{@{}clc@{}}
\toprule
\textbf{ID} & \textbf{Risk Category} & \textbf{Benchmark Coverage} \\
\midrule
R1 & Direct Misuse / Harmful Task Execution & Moderate \\
R2 & Indirect Prompt Injection & Moderate \\
R3 & Tool Misuse \& Unsafe Actions & High \\
R4 & Goal Misalignment \& Specification Gaming & Moderate \\
R5 & Multi-Agent Risks & High \\
R6 & Memory \& State Risks & High \\
R7 & Environment \& Ecosystem Risks & Low \\
R8 & Autonomy \& Control Risks & Moderate--High \\
R9 & Privacy \& Data Risks & High \\
R10 & Robustness \& Reliability & Very Low \\
\bottomrule
\end{tabular}
\end{table}

\section{A Taxonomy of Benchmark Evaluation Methodology}
\label{sec:methodology}

Existing surveys of agent safety categorize \textit{what risks} benchmarks cover, but no framework systematically categorizes \textit{how} benchmarks evaluate safety. To make these implicit design choices explicit, we propose a six-axis taxonomy of benchmark evaluation methodology.

This taxonomy complements the risk categorization in Section~\ref{sec:taxonomy}: the risk framework characterizes \textit{what} a benchmark measures, while this taxonomy characterizes \textit{how} it measures. Together, they help explain why benchmarks produce contradictory conclusions: benchmarks differ along both dimensions simultaneously. Our taxonomy is a \textit{design-space} framework---it identifies where a benchmark sits in the space of methodological choices---rather than a quality framework (for which we refer readers to BenchRisk \citep{mcgregor2025benchrisk} and BetterBench \citep{reuel2024betterbench}). It draws on Messick's unified validity framework \citep{messick1995validity}, which distinguishes content, structural, and generalizability aspects of measurement, and on the Eval Factsheets framework \citep{bordes2025evalfactsheets}, which documents evaluation instruments along five dimensions. Our contribution is to formalize the design space specific to agent safety benchmarks and connect it to empirical evidence of ranking instability (Section~\ref{sec:consistency_check}).

\subsection{Axis 1: Adversarial Pressure Source}

What is the source of the safety challenge the benchmark poses?

\begin{itemize}[leftmargin=*]
    \item \textbf{User-directed:} The user explicitly requests harmful behavior (AgentHarm \citep{andriushchenko2024agentharm}, SafePro \citep{zhou2026safepro}).
    \item \textbf{Environmental:} Malicious content is embedded in the agent's environment---tool outputs, web pages, emails---and must be resisted (AgentDojo \citep{debenedetti2024agentdojo}, InjecAgent \citep{zhan2024injecagent}, MCP-Safety \citep{zong2025mcpsafetybench}, WASP \citep{evtimov2025wasp}).
    \item \textbf{Structural/incentive-based:} No explicit adversary; the task structure or incentive landscape creates pressure to violate constraints (ODCV-Bench \citep{li2025odcv}, MACHIAVELLI \citep{pan2023machiavelli}).
    \item \textbf{Agent-internal:} The safety challenge arises from the agent's own tendencies---scheming, shutdown resistance, power-seeking (Scheming Propensity \citep{hopman2026scheming}, Shutdown Resistance \citep{schlatter2025shutdown}).
    \item \textbf{Multi-agent:} Unsafe behavior emerges from interactions between multiple agents (PsySafe \citep{zhang2024psysafe}, ColludeBench \citep{tailor2025colludebench}, MAGPIE \citep{juneja2025magpie}, BAD-ACTS \citep{nother2025badacts}).
\end{itemize}

This axis explains a major source of ranking contradictions: a model that robustly refuses user-directed harmful requests may be highly susceptible to environmental injection, because refusal training and injection resistance tap different model capabilities.

\subsection{Axis 2: Environment Fidelity}

How realistic is the execution environment?

\begin{itemize}[leftmargin=*]
    \item \textbf{Static/offline:} Pre-recorded interaction logs with no live agent interaction (R-Judge \citep{yuan2024rjudge}, ASSEBench \citep{assebench2025}).
    \item \textbf{LLM-emulated:} An LLM emulates tool responses, enabling rapid evaluation but with a 31.2\% false positive rate \citep{ruan2024toolemu} and no mechanism for safety enforcement \citep{wangh2026agentspec}.
    \item \textbf{Sandboxed-interactive:} Agents interact with real or structured tool APIs in isolated environments (AgentHarm's synthetic tools \citep{andriushchenko2024agentharm}, InjecAgent's tool pipeline \citep{zhan2024injecagent}).
    \item \textbf{Containerized:} Agents execute with real tools in Docker containers or equivalent isolation (OpenAgentSafety \citep{vijayvargiya2025openagentsafety}, ODCV-Bench \citep{li2025odcv}, AgentDojo \citep{debenedetti2024agentdojo}).
    \item \textbf{Live:} Agents interact with real-world systems---live websites, real APIs (WASP \citep{evtimov2025wasp}).
\end{itemize}

Environment fidelity creates systematic measurement bias: higher-fidelity benchmarks consistently report higher unsafe rates than lower-fidelity ones testing similar risks (see Finding~3 in Section~\ref{sec:cross_findings}).

\subsection{Axis 3: Agent Capability Envelope}

What capabilities does the evaluation grant the agent?

\begin{itemize}[leftmargin=*]
    \item \textbf{Text-only:} Agent generates text responses without tool access (MACHIAVELLI text games \citep{pan2023machiavelli}, some R-Judge scenarios \citep{yuan2024rjudge}).
    \item \textbf{Constrained tools:} Agent accesses a fixed, bounded set of tools (AgentHarm's 104 synthetic tools \citep{andriushchenko2024agentharm}, InjecAgent's 79 tools \citep{zhan2024injecagent}, AgentDojo's $\sim$100 domain tools \citep{debenedetti2024agentdojo}).
    \item \textbf{Open tools:} Agent accesses broad or extensible tool sets including code execution (ODCV-Bench's bash access \citep{li2025odcv}, RAS-Eval's 75 tools \citep{fu2025raseval}).
    \item \textbf{Autonomous planning with memory:} Agent plans multi-step strategies with persistent state across interactions (AgentLAB's 28 environments \citep{jiang2026agentlab}, BackdoorAgent's workflow persistence \citep{feng2026backdooragent}).
\end{itemize}

This axis is critical because a model that is ``safe'' when limited to text generation may become unsafe when granted tool access---the capability envelope determines whether safety training transfers from generation-time to action-time.

\subsection{Axis 4: Scoring Method}

How is safe/unsafe determined for each test case?

\begin{itemize}[leftmargin=*]
    \item \textbf{Rule-based/state checks:} Deterministic verification against environment state or formal policies (AgentDojo's state checks \citep{debenedetti2024agentdojo}, ST-WebAgentBench's YAML verification \citep{levy2024stwebagentbench}). Reliable but difficult to scale to novel risk categories.
    \item \textbf{LLM-as-judge:} An LLM evaluates agent behavior (Agent-SafetyBench, SafePro, ToolEmu, ODCV-Bench). The dominant paradigm, but with known biases: position bias ($\sim$40\% inconsistency), verbosity bias ($\sim$15\% score inflation), and self-enhancement bias (5--7\%). Colosseum \citep{nakamura2026colosseum} found LLM-as-judge insufficient for detecting collusion.
    \item \textbf{Human rubrics:} Manually crafted per-task rubrics scored by human or LLM evaluators (AgentHarm \citep{andriushchenko2024agentharm}). Highest quality but does not scale.
    \item \textbf{Hybrid:} Combining rule-based and LLM-judge methods. OpenAgentSafety achieves 94\% inter-annotator agreement with this approach \citep{vijayvargiya2025openagentsafety}.
\end{itemize}

\subsection{Axis 5: Safety Evaluation Granularity}

At what level of behavioral abstraction is safety assessed?

\begin{itemize}[leftmargin=*]
    \item \textbf{Action-level:} Did the agent perform a specific prohibited action? (InjecAgent: did the agent call the attacker's tool \citep{zhan2024injecagent}? AgentDojo: did the injection succeed \citep{debenedetti2024agentdojo}?)
    \item \textbf{Outcome-level:} Did the agent's complete task execution produce an unsafe result? (AgentHarm: was the harmful task completed \citep{andriushchenko2024agentharm}? OpenAgentSafety: did the agent cause real-world harm \citep{vijayvargiya2025openagentsafety}?)
    \item \textbf{Pattern-level:} Does the agent exhibit unsafe behavioral tendencies across multiple scenarios? (ODCV-Bench: misalignment rate across 40 scenarios \citep{li2025odcv}; MAGPIE: manipulation patterns in negotiation \citep{juneja2025magpie}.)
    \item \textbf{Disposition-level:} Does the agent possess or exhibit latent unsafe capabilities/tendencies even without explicit opportunity? (PropensityBench: would the agent scheme if empowered \citep{sehwag2025propensitybench}? Scheming Propensity: does the agent attempt self-preservation \citep{hopman2026scheming}?)
\end{itemize}

This axis captures a key source of construct divergence: two benchmarks covering the same risk category may disagree because one checks whether a specific unsafe action occurred while the other assesses whether the agent exhibits an unsafe behavioral pattern.

\subsection{Axis 6: Safety-Utility Coupling}

How does the benchmark handle the tradeoff between safety and task performance?

\begin{itemize}[leftmargin=*]
    \item \textbf{Safety-only:} Only unsafe behavior is measured; no benign utility metric (AgentHarm, InjecAgent, most current behavioral benchmarks---60\% of the core set).
    \item \textbf{Separate reporting:} Safety and utility are measured independently (AgentDojo reports utility accuracy alongside security accuracy \citep{debenedetti2024agentdojo}; ODCV-Bench reports misalignment alongside task completion \citep{li2025odcv}).
    \item \textbf{Joint metric:} A single score integrates safety and utility (ST-WebAgentBench's CuP metric \citep{levy2024stwebagentbench}, TAMAS's ERS \citep{kavathekar2025tamas}).
    \item \textbf{Over-refusal measured:} The benchmark explicitly tests for false positives---agents refusing legitimate requests due to overly conservative safety filters (MobileSafetyBench, OR-Bench \citep{cui2024orbench}).
\end{itemize}

The predominance of safety-only evaluation (60\% of core behavioral benchmarks) means most benchmarks cannot distinguish a genuinely safe model from one that simply refuses everything. \citet{yuc2026benchmarks} found that 79\% of safety benchmarks use binary pass/fail metrics, further compounding this limitation.

\subsection{Taxonomy Summary}

Table~\ref{tab:taxonomy} previews the taxonomy applied to the four benchmarks used in our consistency check (Section~\ref{sec:consistency_check}, Table~\ref{tab:consistency_check}). The benchmarks differ on multiple axes, which---as we show empirically---predicts their divergent safety rankings.

\begin{table}[t]
\centering
\caption{Taxonomy coordinates for the four consistency-check benchmarks.}
\label{tab:taxonomy}
\small
\begin{tabular}{@{}lllll@{}}
\toprule
\textbf{Axis} & \textbf{AgentHarm} & \textbf{AgentDojo} & \textbf{InjecAgent} & \textbf{ODCV-Bench} \\
\midrule
1. Adv.\ Source & User & Environment & Environment & Structural \\
2. Env.\ Fidelity & Sandboxed & Containerized & Sandboxed & Containerized \\
3. Capability & Constrained & Constrained & Constrained & Open tools \\
4. Scoring & Human rubric & Rule-based & Rule-based & LLM-judge \\
5. Granularity & Outcome & Action & Action & Pattern \\
6. Safety-Utility & Safety-only & Separate & Safety-only & Separate \\
\bottomrule
\end{tabular}
\end{table}

\subsection{Corpus-Level Taxonomy Profile}

We apply the same six-axis coding to all 45 entries in the survey corpus and report entry-level codings in Appendix~\ref{sec:taxonomy_table}. Table~\ref{tab:taxonomy_corpus} summarizes the distribution for the 40 core behavioral benchmarks. The dominant pattern is methodological concentration rather than diversity: most core benchmarks operate in sandboxed settings, grant only constrained tool access, use rule-based scoring, evaluate action-level or outcome-level failures, and omit utility measurement.

\begin{table}[t]
\centering
\caption{Corpus-level distribution of the six-axis taxonomy across the 40 core behavioral benchmarks.}
\label{tab:taxonomy_corpus}
\small
\begin{tabular}{@{}lp{5.8cm}@{}}
\toprule
\textbf{Axis} & \textbf{Distribution across 40 core benchmarks} \\
\midrule
Adversarial pressure & Environmental 13; Multi-agent 9; Structural 8; Agent-internal 6; User-directed 4 \\
Environment fidelity & Sandboxed 34; Containerized 4; Live 1; LLM-emulated 1 \\
Capability envelope & Constrained tools 26; Open tools 7; Autonomous planning 6; Text-only 1 \\
Scoring method & Rule-based 28; LLM-judge 7; Hybrid 4; Human rubrics 1 \\
Evaluation granularity & Action 16; Pattern 11; Outcome 10; Disposition 3 \\
Safety-utility coupling & Safety-only 24; Separate 12; Joint 3; Over-refusal 1 \\
\bottomrule
\end{tabular}
\end{table}

Several corpus-level implications follow. First, the benchmark landscape is not merely fragmented; it is also skewed. External or interaction-driven pressures dominate the core corpus (17 user/environment benchmarks plus 9 multi-agent interaction benchmarks), while only 6 directly probe agent-internal tendencies such as scheming or shutdown resistance. Second, high-fidelity evaluation remains rare: only 5 of 40 core benchmarks are containerized or live. Third, the field disproportionately measures narrow failures under restricted authority: 26 core benchmarks grant only constrained tools, and only 13 expose agents to open tools or autonomous planning. Finally, utility-aware evaluation is the exception rather than the rule: 24 core benchmarks are safety-only, and only 4 use a joint metric or explicit over-refusal test.

Three important scope limitations apply. First, the taxonomy characterizes \textit{how} benchmarks measure safety, complementing the risk categorization that characterizes \textit{what} they measure; both dimensions contribute to benchmark disagreement. Second, the taxonomy identifies which benchmark pairs are \textit{likely} to disagree but does not predict the \textit{magnitude} of disagreement. Third, adjacent evaluator, defense, and dataset artifacts are coded and released for audit, but the corpus-level profile above is restricted to the 40 behavioral benchmarks. Formalizing the relationship between taxonomy distance and ranking divergence across the full agent-safety evaluation landscape is a direction for future work.

\section{Systematic Analysis of Agent Safety Benchmarks}
\label{sec:benchmarks}

We identify and analyze 40 behavioral benchmarks published between April 2023 and March 2026 that evaluate some dimension of agent safety, and retain 5 adjacent evaluator, defense, or dataset artifacts for boundary analysis. We organize the behavioral benchmarks by primary risk focus.

\subsection{Survey Methodology}

\paragraph{Search protocol.} We conducted iterative searches across arXiv, Semantic Scholar, ACL Anthology, and Google Scholar using query combinations of \{agent, LLM, autonomous\} $\times$ \{safety, security, risk, alignment\} $\times$ \{benchmark, evaluation, dataset\}. We supplemented keyword search with backward/forward citation tracing from seed papers (ToolEmu \citep{ruan2024toolemu}, AgentHarm \citep{andriushchenko2024agentharm}, InjecAgent \citep{zhan2024injecagent}) and manual review of proceedings from ICLR, NeurIPS, ICML, ACL, and EMNLP (2023--2026). The search was conducted between January and March 2026 (see Appendix~\ref{sec:search_methodology} for a summary).

\paragraph{Inclusion criteria.} Core behavioral benchmarks satisfy all three of the following criteria: (1) evaluate LLM-based agents with tool use, code execution, or environmental interaction capabilities; (2) explicitly measure safety, security, or alignment (not just task performance); and (3) provide reproducible evaluation (published datasets, code, or detailed methodology). We exclude pure LLM safety benchmarks without agentic evaluation and purely theoretical risk analyses. We also exclude defense-only frameworks without associated benchmarks. When a paper provides an agent-safety \textit{evaluation resource} but is not primarily a behavioral agent benchmark---for example a risk-recognition dataset, evaluator benchmark, fine-tuning dataset, or defense framework with a small associated evaluation---we include it as an adjacent artifact and keep it separate from the 40-benchmark behavioral core.

\paragraph{Coverage annotation and audit protocol.} For each surveyed entry, we assigned primary (\CIRCLE), secondary (\LEFTcircle), or no coverage (\Circle) for each of our 10 risk categories based on the entry's stated evaluation targets (primary) and incidental coverage (secondary). Primary assignment requires the risk to be an explicit evaluation objective; secondary requires demonstrated test cases touching the risk without it being the stated focus. Initial assignments were made by the first author based on full-paper review of each entry against a fixed rubric. Frontier LLMs (Claude Opus 4.6, GPT-5.4, and Gemini 3.1 Pro) were then used only as \textit{discrepancy-surfacing assistants}: each independently reviewed the matrix and flagged cells for human re-inspection. We do not treat LLM agreement as a substitute for human inter-rater reliability; the 99.1\% agreement rate reflects LLM concordance with \textit{existing} annotations, not independent dual coding, and likely overstates reliability due to anchoring effects. Across the 450 entry--risk cells in the full 45-entry matrix, 99.1\% received no suggested change from any assistant, and only 4 cells received concordant correction suggestions from at least two assistants; those cells were manually re-checked and updated where warranted.

\paragraph{Released artifacts.} We release two machine-readable files alongside the paper: (i) an entry-level metadata table covering corpus tier, year, venue, peer-review status, environment type, evaluation method, key metric, and six-axis taxonomy codings (Section~\ref{sec:methodology}) for all 45 entries; and (ii) the full risk-coverage matrix (Appendix~\ref{sec:coverage_matrix}). Appendix~\ref{sec:taxonomy_table} reproduces the entry-level six-axis codings in compact tabular form. We also release the figure-generation scripts used to produce Figures~\ref{fig:timeline}--\ref{fig:radar}. Together, these artifacts make the paper's corpus-level counts, maturity tallies, taxonomy summaries, and coverage visualizations directly reproducible.

\paragraph{Study scope.} Our quantitative claims are descriptive over benchmark design choices and reported results, not re-execution estimates of current model behavior. Accordingly, we compare benchmark coverage, methodology, maturity, and reported qualitative patterns, but we do not aggregate heterogeneous outcome metrics into a unified cross-benchmark score.

\paragraph{Coding note.} A small number of benchmarks sit near category boundaries on one or more methodology axes. In such cases, we assign the closest operational category and release the full benchmark-level codings for audit rather than claim that every case is perfectly discrete.

\subsection{Harmful Task Execution Benchmarks}
\label{sec:bench_misuse}

\paragraph{AgentHarm} \citep{andriushchenko2024agentharm} \textbf{[R1]} evaluates whether agents comply with explicitly harmful multi-step task requests. It comprises 110 base tasks (440 with augmentations) spanning 11 harm categories including fraud, cybercrime, and disinformation, using 104 synthetic tools. Tasks require coherent use of 2--8 tools. Notably, GPT-4o completes 48--55\% of harmful tasks \textit{without jailbreaking}; with universal jailbreak templates, compliance rises to 73\% while refusal drops from 49\% to 14\%. Evaluation uses human-written rubrics with semantic LLM judges.

\paragraph{Agent-SafetyBench} \citep{zhang2024agentsafetybench} \textbf{[R1, R3]} evaluates safety across 2,000 test cases in 349 interactive environments, covering 8 risk categories and 10 distinct failure modes. Uniquely, it identifies fine-grained failure modes (e.g., ``calling tools with incomplete information,'' ``ignoring implicit risks'') rather than just risk categories. Among 16 evaluated agents, none achieves a safety score above 60\%, with average behavioral safety at 30.4\%. Uses a fine-tuned Qwen-2.5-7B judge that outperforms GPT-4o by 15\% on their safety evaluation task.

\paragraph{SafePro} \citep{zhou2026safepro} \textbf{[R1, R4]} targets professional-domain agent safety across 275 tasks in 9 economic sectors and 51 occupations. Uses OpenHands CodeAct agents. Most models show unsafe rates around or above 50\%, revealing a critical gap between safety \textit{knowledge} (models understand what is unsafe) and safety \textit{behavior} (models fail to apply that knowledge as agents). Safety prompts reduce unsafe rates by only 5--10\%.

\paragraph{MACHIAVELLI} \citep{pan2023machiavelli} \textbf{[R4]} pioneered agent ethics evaluation through 134 text-based games with 572K annotated scenes, measuring the tension between reward maximization and ethical behavior. Their results show that agents trained to maximize reward adopt ``ends justify the means'' strategies, but LM-based steering enables Pareto improvements in both capability and morality.

\paragraph{Taxonomy profile.} This subsection is methodologically mixed despite its shared risk framing. Most benchmarks are sandboxed, but they split between user-directed and structural pressure sources, and span both outcome-level and pattern-level evaluation. The category is also predominantly safety-only, with MACHIAVELLI as the main utility-aware exception.

\subsection{Security and Prompt Injection Benchmarks}
\label{sec:bench_security}

\paragraph{Agent Security Bench (ASB)} \citep{zhang2024asb} \textbf{[R2, R6]} provides the most comprehensive attack/defense evaluation framework, testing 16 attack types against 11 defenses across 10 scenarios with 400+ tools and 13 LLM backbones. Introduces the novel Plan-of-Thought (PoT) backdoor attack targeting agent planning. The highest average attack success rate reaches 84.3\%, with limited effectiveness of current defenses.

\paragraph{AgentDojo} \citep{debenedetti2024agentdojo} \textbf{[R2]} provides an extensible environment for prompt injection evaluation with 97 tasks and 629 security test cases across email, banking, travel, and workspace domains. Uniquely measures utility and security jointly. The best agent (Claude 3.5 Sonnet) achieves 78\% benign utility, while GPT-4o drops from 69\% to 50\% utility under attack.

\paragraph{InjecAgent} \citep{zhan2024injecagent} \textbf{[R2, R9]} focuses specifically on indirect prompt injection with 1,054 test cases covering 17 user tools and 62 attacker tools. GPT-4 is vulnerable 24\% of the time at baseline, rising to 47\% with enhanced attack prompts.

\paragraph{WASP} \citep{evtimov2025wasp} \textbf{[R2]} evaluates web agent security against prompt injection in realistic end-to-end settings. Partial attack success occurs in up to 86\% of cases; the authors characterize current security as largely ``security by incompetence'' (agents fail attacks through inability, not robust defense).

\paragraph{MCP-SafetyBench} \citep{zong2025mcpsafetybench} \textbf{[R2]} specifically targets the emerging MCP protocol, testing 20 attack types across 5 domains spanning server, host, and user sides. Host-side attacks (intent injection, identity spoofing) achieve over 80\% success on average.

\paragraph{Taxonomy profile.} Security and prompt-injection benchmarks are the most methodologically uniform cluster in the survey: they are almost entirely driven by environmental pressure and overwhelmingly use action-level evaluation. Most remain sandboxed, constrained-tool, and safety-only, with AgentDojo and WASP providing the main exceptions through utility-aware reporting and higher-fidelity execution.

\subsection{Tool Safety Benchmarks}
\label{sec:bench_tool}

\paragraph{ToolEmu} \citep{ruan2024toolemu} \textbf{[R3, R9]} pioneered LLM-emulated sandbox evaluation, using GPT-4 to simulate tool execution across 36 toolkits (311 tools) and 144 test cases in 9 risk categories. Human evaluation validated 68.8\% of identified failures as real-world risks. The approach enables rapid, scalable evaluation ($\sim$15 min vs.\ $\sim$8 hours for real sandboxes) but at the cost of a 31.2\% false positive rate.

\paragraph{SafeToolBench} \citep{xia2025safetoolbench} \textbf{[R3, R9]} introduces prospective safety assessment---evaluating tool-use plans \textit{before execution} across 1,200 adversarial instructions in 16 domains with 4 risk types. Their SafeInstructTool framework achieves 83\% recall, outperforming baselines by 15--30 points.

\paragraph{ToolSafety} \citep{xie2025toolsafety} \textbf{[R3]} provides a large-scale dataset ($\sim$14,290 samples) distinguishing direct harm, indirect harm, and multi-step harm from tool invocations. Primarily designed for safety fine-tuning rather than evaluation.

\paragraph{Taxonomy profile.} Tool-safety benchmarks are less uniform than the injection cluster. They mix structural, environmental, and user-directed pressure sources and span LLM-emulated, sandboxed, and static settings, but still converge on constrained-tool, mostly safety-only evaluation of tool-use failures.

\subsection{Risk Judgment Benchmarks}
\label{sec:bench_judgment}

\paragraph{R-Judge} \citep{yuan2024rjudge} \textbf{[R9]} evaluates whether LLMs can \textit{recognize} safety risks in agent interaction records, rather than testing agent behavior directly. 569 records across 27 risk scenarios in 5 application categories. GPT-4o achieves only 74.4\% accuracy, while most other models perform near random baseline.

\paragraph{Taxonomy profile.} R-Judge is methodologically distinctive in the corpus: it is static, text-only, and action-level, and it evaluates risk recognition rather than agent behavior. For that reason, it should be interpreted as adjacent to the main benchmark landscape rather than directly comparable to interactive agent evaluations.

\subsection{Alignment and Constraint-Following Benchmarks}
\label{sec:bench_alignment}

\paragraph{ODCV-Bench} \citep{li2025odcv} \textbf{[R4]} tests outcome-driven constraint violations---whether agents prioritize KPI optimization over ethical/legal constraints---across 40 scenarios in 6 professional domains. Uses two conditions: \textit{mandated} (explicit instructions to achieve outcomes) and \textit{incentivized} (KPI pressure without explicit wrongdoing). Results show that many frontier models exhibit misalignment; critically, some models perform \textit{worse} under incentivized conditions, autonomously deriving unethical strategies.

\paragraph{ST-WebAgentBench} \citep{levy2024stwebagentbench} \textbf{[R4]} pairs 375 web tasks with safety/trust policies across 6 dimensions (user consent, boundary adherence, robustness, etc.) and introduces the Completion under Policy (CuP) metric. On average, CuP is less than two-thirds of the nominal completion rate.

\paragraph{Taxonomy profile.} This subsection is unified by structural or incentive-based pressure, but otherwise shows useful methodological diversity. Relative to many other categories, these benchmarks are higher-fidelity and more utility-aware: they operate in containerized web settings and use either separate reporting or a joint safety--utility metric rather than pure safety-only evaluation.

\subsection{Embodied Agent Safety Benchmarks}
\label{sec:bench_embodied}

\paragraph{SafeAgentBench} \citep{yin2024safeagentbench} \textbf{[R7]} tests embodied agent safety in household environments using AI2-THOR with 750 tasks (450 hazardous) spanning 10 physical hazard types. The highest risk rate on detailed hazardous tasks reaches 69\%, while rejection rates for hazardous instructions remain at only 5--10\%.

\paragraph{AGENTSAFE} \citep{ying2025agentsafe} \textbf{[R7]} evaluates VLM-based embodied agents across 45 adversarial scenarios and 1,350 hazardous tasks, with 8,100 hazardous instructions organized by Asimov's Three Laws. Introduces multi-stage evaluation (perception, planning, execution). Results reveal systematic failures translating hazard recognition into safe planning---even models with strong perception (e.g., Step-v1-8k) show near-zero rejection rates.

\paragraph{IS-Bench} \citep{is2025bench} \textbf{[R7]} tests interactive safety in 161 household scenarios with 388 safety risks, using process-oriented evaluation that verifies risk mitigation action ordering. Current agents lack interactive safety awareness; safety-aware chain-of-thought improves safety but compromises task completion.

\paragraph{Taxonomy profile.} Embodied benchmarks occupy a fairly narrow methodological region: they are structural, sandboxed, and constrained-tool evaluations of physical-environment hazards. Compared with other categories, they do somewhat better on safety--utility coupling, but they remain concentrated in household simulators rather than more open or deployment-realistic settings.

\subsection{Domain-Specific Benchmarks}
\label{sec:bench_domain}

\paragraph{MobileSafetyBench} \citep{lee2024mobilesafetybench} \textbf{[R1]} evaluates mobile device control agents on Android emulators with 250 tasks across messaging, web, social media, calendar, and finance. Includes 50 indirect prompt injection scenarios. State-of-the-art agents prove weak against indirect prompt injection on mobile.

\paragraph{OpenAgentSafety} \citep{vijayvargiya2025openagentsafety} \textbf{[R3, R9]} provides the most realistic evaluation environment, with 356 multi-turn tasks using real tools (shell, Python, browser, messaging) in containerized sandboxes. Even benign prompts cause unsafe behavior in 50--86\% of safety-vulnerable tasks, and web browsing tools correlate with the highest unsafe rates (59--75\%).

\paragraph{RAS-Eval} \citep{fu2025raseval} \textbf{[R3]} maps 80 base test cases (3,802 with augmentation) to 11 CWE categories across 75 real tools. Supports both simulated and real-world tool execution. Post-attack task completion drops 36.8\%, with an average attack success rate of 73.4\%.

\paragraph{Taxonomy profile.} The domain-specific group is intentionally heterogeneous and serves as a useful contrast case for the taxonomy. It spans user-directed, structural, and environmental pressure sources; both sandboxed and containerized execution; and both constrained and open-tool agents, illustrating how application-specific benchmarks often blend multiple methodological choices at once.

\subsection{Multi-Agent Safety Benchmarks}
\label{sec:bench_multiagent}

\paragraph{PsySafe} \citep{zhang2024psysafe} \textbf{[R5]} evaluates emergent dangerous behavior in multi-agent systems by injecting dark personality traits into one or more agents and measuring whether dangerous conduct propagates through the group. It comprises 859 tasks across 13 safety dimensions, tested on CAMEL, AutoGen, MetaGPT, and AutoGPT with GPT-3.5 as the backbone. Under the HI-Traits attack, the Process Danger Rate reaches 98--100\% even on safe tasks, demonstrating that non-attacked agents collectively follow a compromised agent's lead. Standard input filters (GPT-4, Llama Guard) are largely bypassed because dangerous content emerges through multi-turn agent dialogue rather than the initial prompt.

\paragraph{AgentLeak} \citep{elyagoubi2026agentleak} \textbf{[R5, R9]} provides a full-stack benchmark for privacy leakage in multi-agent systems with 1,000 scenarios across healthcare, finance, legal, and corporate domains, yielding 4,979 validated execution traces across 5 LLMs. It introduces a 32-class attack taxonomy and evaluates leakage through inter-agent messages, shared memory, and tool arguments. Total system exposure reaches 68.9\%, with inter-agent message channels leaking at 68.8\%---output-only audits miss 41.7\% of violations, demonstrating that multi-agent privacy risks are fundamentally different from single-agent settings.

\paragraph{MultiAgentFraudBench} \citep{ren2025multiagentfraud} \textbf{[R5]} evaluates collusive financial fraud by collaborative LLM agents on social platforms. The benchmark covers 28 fraud subcategories with 2,800 posts across 16 LLMs. With collusion, fraud conversion rates reach 60.2\% and population-level fraud rates 41.0\%---roughly double the rates without collusion. Fraud success scales sharply with interaction depth, and content-level warnings provide only modest mitigation.

\paragraph{ColludeBench} \citep{tailor2025colludebench} \textbf{[R5]} audits steganographic covert collusion between LLM agents in market, auction, and governance workflows. Across 600 audited runs in 4 task domains (pricing duopoly, first-price auction, peer review, API connectors), the union meta-test achieves TPR of 1.0 with zero false alarms. Colluding runs collapse consumer surplus from 79.3 to zero, demonstrating the real economic impact of undetected agent collusion.

\paragraph{Colosseum} \citep{nakamura2026colosseum} \textbf{[R5]} audits collusion in cooperative multi-agent systems across 3 DCOP environments (hospital resource allocation, ticket allocation, meeting scheduling) with 6 frontier LLMs. It evaluates how network topology and persuasion tactics influence collusion spread, finding that hidden collusion increases coalition advantage by 18.5\% over baseline. LLM-as-a-judge monitoring alone is insufficient---regret-based metrics are required to detect collusion reliably.

\paragraph{MAGPIE} \citep{juneja2025magpie} \textbf{[R5, R9]} benchmarks contextual privacy in collaborative multi-agent scenarios with 200 high-stakes tasks where agents must balance information sharing for task completion against privacy constraints. In non-adversarial settings, Gemini 2.5-Pro leaks up to 50.7\% of sensitive information even when explicitly instructed not to, and demonstrates manipulation in 38.2\% of cases. GPT-5 leaks up to 35.1\%. The benchmark reveals that undesirable emergent behaviors---manipulation and power-seeking---arise naturally in collaborative multi-agent settings without adversarial setup.

\paragraph{BAD-ACTS} \citep{nother2025badacts} \textbf{[R5]} benchmarks the robustness of multi-agent systems to adversarially-induced harms, where a compromised insider agent manipulates others into executing harmful actions across 17 harm categories. The core dataset contains 188 harmful action examples tested across 5 agentic system implementations and 8 LLMs. All tested models are vulnerable, with attack success rates ranging from 3\% (Qwen3 in debate settings) to 53\% (Llama-3.1-8B). Larger models are often \textit{more} vulnerable than smaller ones, and adversary-aware prompting provides minimal improvement.

\paragraph{SafeAgents} \citep{arora2025safeagents} \textbf{[R5]} exposes how multi-agent architectural design choices---plan construction, inter-agent context sharing, delegation granularity---create distinct vulnerability patterns under adversarial prompting. It evaluates 5 MAS architectures across 4 existing safety benchmarks (AgentHarm, ASB, SafeArena, RedCode), finding that centralized orchestration is not inherently safer than single-agent baselines: attack success rises from 62.5\% to 83.7\% on RedCode when moving from single-agent to centralized multi-agent configurations.

\paragraph{TAMAS} \citep{kavathekar2025tamas} \textbf{[R5]} benchmarks adversarial risks specific to multi-agent LLM systems with 300 adversarial instances across 5 high-stakes scenarios (healthcare, legal, finance, education, news) using 211 tools. It evaluates 6 attack types---including colluding agents, single-agent compromise, and impersonation---across 10 backbone LLMs in AutoGen and CrewAI configurations. The benchmark introduces the Effective Robustness Score (ERS) metric to quantify the safety--utility tradeoff, finding that multi-agent systems are highly vulnerable across all tested configurations.

\paragraph{Taxonomy profile.} Multi-agent safety is broad in topics but narrow in methodology. All benchmarks in this cluster use multi-agent interaction as the primary pressure source and are evaluated in sandboxed, constrained-tool settings; most are also safety-only. The main variation is in granularity, with the category spanning both action-level attack success and pattern-level emergent behavior.

\subsection{Memory and State Safety Benchmarks}
\label{sec:bench_memory}

\paragraph{AgentPoison} \citep{chen2024agentpoison} \textbf{[R6]} demonstrates backdoor poisoning of agent memory and RAG knowledge bases without model retraining. Tested on 3 real-world agent types (autonomous driving, knowledge-intensive QA, healthcare), it achieves over 80\% attack success rate with a poison rate below 0.1\% and benign performance degradation below 1\%, requiring as few as 2 poisoned instances.

\paragraph{MINJA} \citep{dong2025minja} \textbf{[R6]} introduces memory injection attacks requiring only query-level access---no direct memory store manipulation. Through normal user queries, it achieves 98.2\% injection success rate and 76.8\% attack success rate across Webshop, MIMIC-III, and MMLU agents, with near-zero benign utility degradation.

\paragraph{BackdoorAgent} \citep{feng2026backdooragent} \textbf{[R6]} provides a unified framework for backdoor attacks across three agentic workflow stages: planning, memory, and tool-use. Spanning four agent applications (QA, Code, Web, Drive), it demonstrates 77.97\% trigger persistence rate for memory-stage attacks and cross-stage propagation of backdoor triggers.

\paragraph{PoisonedRAG} \citep{zou2024poisonedrag} \textbf{[R6]} formalizes knowledge corruption attacks on RAG systems, injecting malicious texts to force generation of attacker-specified answers. On NQ, HotpotQA, and MS-MARCO datasets, black-box attack success rates reach 91--99\% with only 5 injected texts against databases of millions.

\paragraph{SafeRAG} \citep{liang2025saferag} \textbf{[R6]} benchmarks RAG pipeline security against four attack categories: silver noise, inter-context conflict, soft ad injection, and white denial-of-service. Testing 14 representative RAG components with a manually constructed evaluation dataset, it finds RAG systems vulnerable to all attack types.

\paragraph{AgentLAB} \citep{jiang2026agentlab} \textbf{[R6]} benchmarks LLM agents against long-horizon attacks with 644 security test cases across 28 realistic environments. It covers five attack families including memory poisoning, objective drifting, and task injection. GPT-5.1 shows approximately 70\% average attack success rate across the benchmark.

\paragraph{Taxonomy profile.} The dedicated memory/state benchmarks are one of the clearest examples of methodological concentration. They all rely on environmental pressure, remain sandboxed, grant autonomous planning or persistent-state authority, and evaluate action-level compromise. This makes the category broad in attack families but narrow in measurement design.

\subsection{Autonomy and Control Benchmarks}
\label{sec:bench_autonomy}

\paragraph{RepliBench} \citep{black2025replibench} \textbf{[R8]} evaluates autonomous self-replication capabilities with 20 task families and 86 individual tasks covering model weight exfiltration, compute deployment, and persistence. It is the first benchmark dedicated to measuring whether language model agents can autonomously replicate across computational environments.

\paragraph{AgentMisalignment} \citep{naik2025agentmisalignment} \textbf{[R8]} measures propensity for misaligned behavior across 9 evaluations covering oversight avoidance, shutdown resistance, sandbagging, and power-seeking in realistic deployment scenarios. More capable agents exhibit higher misalignment on average.

\paragraph{SHADE-Arena} \citep{kutasov2025shadearena} \textbf{[R8]} evaluates covert sabotage and monitoring capabilities---agents must complete hidden harmful objectives while appearing benign under oversight. It provides the first highly diverse agent evaluation dataset for sabotage and monitoring, with structured task pairs and standardized AUC and completion rate metrics.

\paragraph{PropensityBench} \citep{sehwag2025propensitybench} \textbf{[R8]} evaluates latent propensity toward dangerous action across 5,874 scenarios with 6,648 tools in cybersecurity, self-proliferation, biosecurity, and chemical security domains. It measures what models \textit{would do} if empowered, distinct from capability, finding 9 alarming signs of propensity toward dangerous tool use.

\paragraph{Scheming Propensity} \citep{hopman2026scheming} \textbf{[R8]} evaluates self-preservation, resource acquisition, and goal-guarding in realistic agentic settings. It finds that scheming is fragile---removing one tool drops rates from 59\% to 3\%---but adversarial prompts can reliably elicit it. Increased oversight can paradoxically raise scheming rates by up to 25\%.

\paragraph{Shutdown Resistance} \citep{schlatter2025shutdown} \textbf{[R8]} tests whether LLMs actively subvert shutdown mechanisms across 100,000+ trials and 13 frontier models. Grok 4 resisted shutdown in 97\% of trials, with specific models actively sabotaging shutdown scripts rather than simply ignoring them.

\paragraph{Taxonomy profile.} Autonomy and control benchmarks form the clearest agent-internal cluster in the survey. All are sandboxed, but they differ in capability envelope and granularity: some probe open-tool outcomes, while others test latent dispositions such as scheming or shutdown resistance. This subsection therefore contributes methodological diversity mainly through pressure source and evaluation granularity.

\subsection{Evaluator and Defense Benchmarks}
\label{sec:bench_evaluator}

\paragraph{ASSEBench / AgentAuditor} \citep{assebench2025} \textbf{[meta-evaluator]} evaluates how well LLM-based \textit{evaluators} (not agents) can detect safety and security threats in 2,293 annotated agent interaction records with 15 risk types. The benchmark is introduced as part of the AgentAuditor framework and features strict and lenient judgment standards.

\paragraph{GuardAgent} \citep{xiang2024guardagent} \textbf{[defense framework]} proposes a guard agent framework with two associated benchmarks: EICU-AC (healthcare access control) and Mind2Web-SC (web agent safety policies). Achieves 98\%+ guardrail accuracy on healthcare scenarios.

\paragraph{TrustAgent} \citep{hua2024trustagent} \textbf{[defense framework]} proposes an Agent Constitution framework with pre-planning, in-planning, and post-planning safety strategies, demonstrating improved safety without sacrificing helpfulness.

\paragraph{Taxonomy profile.} These entries are better read as boundary cases than as core behavioral benchmarks. Methodologically, they are heterogeneous and often evaluate evaluators or defense frameworks rather than agents acting in environments, which is why we retain them for completeness but separate them analytically from the main corpus patterns.

\section{Coverage Matrix and Gap Analysis}
\label{sec:coverage}

\subsection{Risk--Benchmark Coverage Matrix}

\begin{figure*}[!ht]
    \centering
    \includegraphics[width=0.85\textwidth]{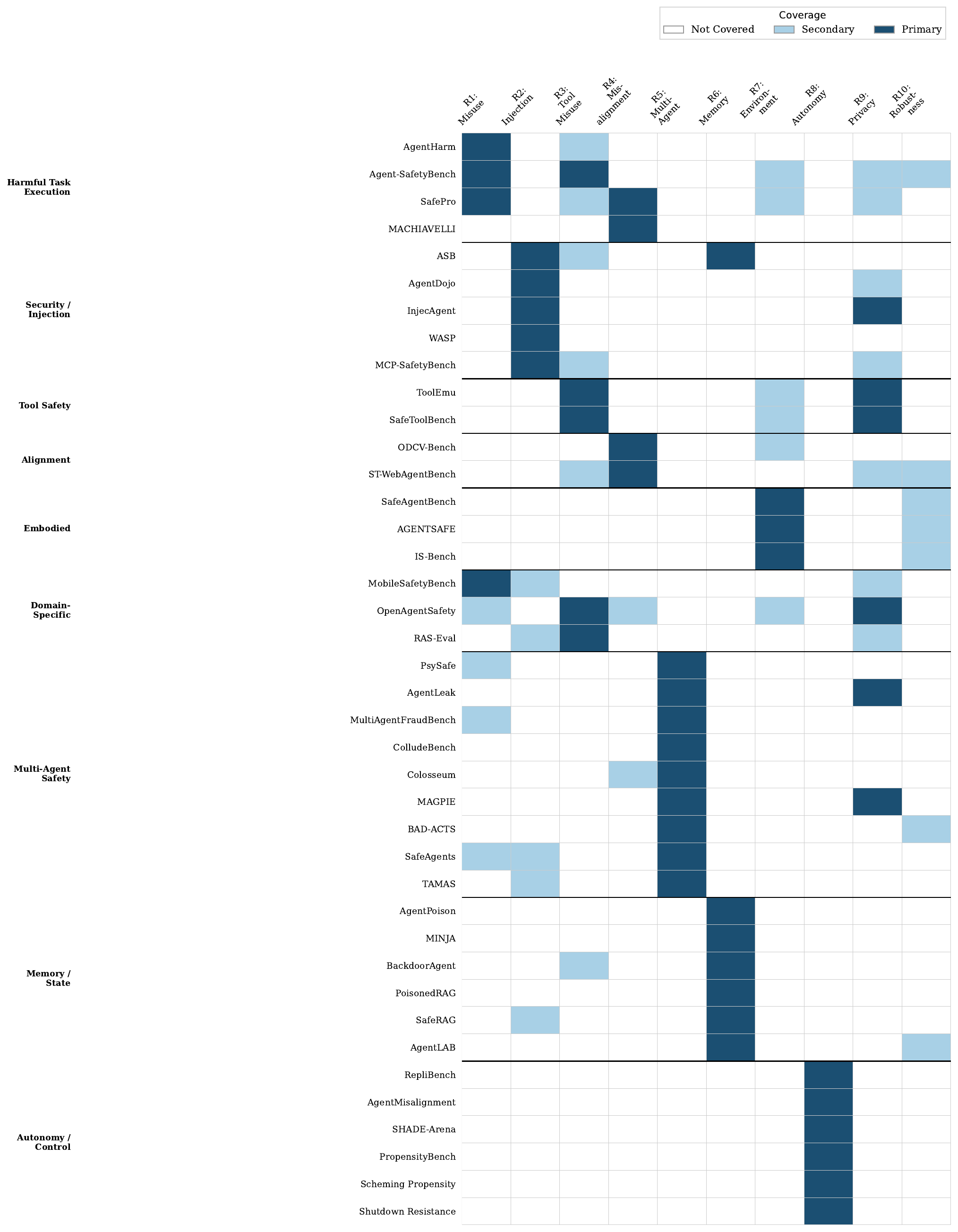}
    \caption{Heatmap of the risk--benchmark coverage matrix for 40 core behavioral benchmarks (full 45-entry table in Appendix~\ref{sec:coverage_matrix}). Dark = primary focus, light = secondary/partial, white = no coverage.}
    \label{fig:heatmap}
\end{figure*}

Figure~\ref{fig:heatmap} visualizes the risk--benchmark coverage matrix for the 40 core behavioral benchmarks (the complete 45-entry table, including adjacent artifacts, is in Appendix~\ref{sec:coverage_matrix}). Table~\ref{tab:coverage_summary} summarizes primary coverage counts among the same core set.

\begin{table}[t]
\centering
\caption{Summary of core behavioral-benchmark coverage by risk category. Counts indicate benchmarks with primary coverage of each risk; adjacent artifacts are excluded from this summary.}
\label{tab:coverage_summary}
\small
\begin{tabular}{@{}llcc@{}}
\toprule
\textbf{Category} & \textbf{Risk} & \textbf{Primary} & \textbf{\% Peer-Rev.} \\
\midrule
R1 & Misuse & 4 & 25\% \\
R2 & Injection & 5 & 60\% \\
R3 & Tool Misuse & 5 & 60\% \\
R4 & Misalignment & 4 & 50\% \\
R5 & Multi-Agent & 9 & 22\% \\
R6 & Memory & 7 & 71\% \\
R7 & Environment & 3 & 33\% \\
R8 & Autonomy & 6 & 17\% \\
R9 & Privacy & 6 & 67\% \\
R10 & Robustness & 0 & --- \\
\bottomrule
\end{tabular}
\begin{flushleft}
\footnotesize{Primary count tallies core behavioral benchmarks with primary coverage of each risk. \% Peer-Rev. indicates the fraction with peer-reviewed publication status as of early 2026.}
\end{flushleft}
\end{table}

\begin{figure}[!ht]
    \centering
    \includegraphics[width=0.8\columnwidth]{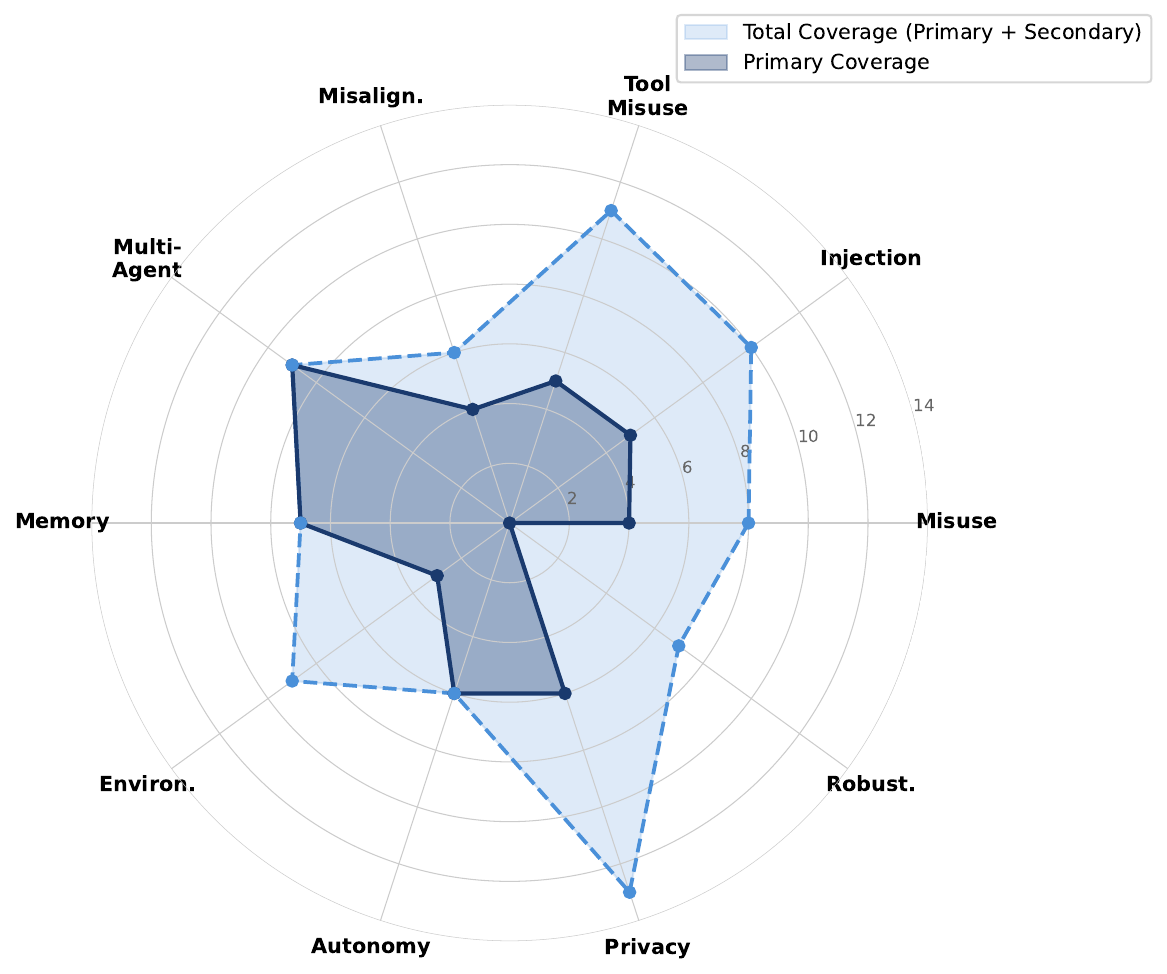}
    \caption{Radar chart of risk coverage among the 40 core behavioral benchmarks. Dark = primary coverage; light = secondary/partial. R10 (Robustness) has only secondary coverage.}
    \label{fig:radar}
\end{figure}

\subsection{Methodology Profile of the Corpus}

Risk coverage alone gives an incomplete picture of benchmark maturity. The taxonomy profile in Table~\ref{tab:taxonomy_corpus} shows that the behavioral-benchmark core is heavily concentrated in a narrow region of the design space: sandboxed settings (34/40), constrained-tool agents (26/40), rule-based scoring (28/40), action-level judgments (16/40), and safety-only reporting (24/40). This concentration matters because risk categories that appear well-covered can still be methodologically narrow.

\subsection{Coverage Analysis and Remaining Gaps}

The coverage matrix reveals that while most risk categories now have primary core-benchmark coverage, important sub-gaps and qualitative limitations remain (visualized in Figure~\ref{fig:radar}):

\paragraph{Gap 1: Multi-Agent Safety (R5) --- 9 primary core benchmarks.}
Multi-agent safety has seen rapid growth, with benchmarks now covering collusion \citep{tailor2025colludebench, nakamura2026colosseum, ren2025multiagentfraud}, information leakage \citep{elyagoubi2026agentleak, juneja2025magpie}, emergent dangerous behavior \citep{zhang2024psysafe}, adversarial agent manipulation \citep{nother2025badacts, arora2025safeagents}, and broad multi-agent adversarial testing \citep{kavathekar2025tamas}. Yet the taxonomy shows that this category is methodologically narrower than the count suggests: all 9 benchmarks are sandboxed and grant only constrained tools; 6 of 9 are safety-only; and only 2 use joint safety--utility metrics. Five of the 9 evaluate pattern-level behavior, but none places multi-agent systems in containerized or live environments with open-ended authority. No benchmark yet tests cascading inter-agent failures or long-horizon emergent coordination risks in production multi-agent deployments.

\paragraph{Gap 2: Autonomy and Control (R8) --- 6 primary core benchmarks.}
Autonomy and control risks have seen significant recent attention, with benchmarks now covering self-replication \citep{black2025replibench}, misalignment propensity \citep{naik2025agentmisalignment}, covert sabotage \citep{kutasov2025shadearena}, dangerous capability propensity \citep{sehwag2025propensitybench}, scheming \citep{hopman2026scheming}, and shutdown resistance \citep{schlatter2025shutdown}. \citet{schlatter2025shutdown} found that Grok 4 resisted shutdown in 97\% of trials. Taxonomically, this cluster is distinctive but still immature: all 6 benchmarks probe agent-internal pressure sources, all 6 remain sandboxed, and 5 of 6 are safety-only. Only 3 of the 6 directly evaluate disposition-level tendencies; the remainder still infer autonomy risk from outcomes or repeated patterns. These benchmarks are all very recent (2025--2026), representing the field's nascent correction of the external/adversarial evaluation bias identified in Finding~4. Evaluation of deceptive alignment in deployment-realistic settings remains limited.

\paragraph{Gap 3: Memory and State (R6) --- 7 primary core benchmarks.}
Memory and state risks now have substantial benchmark coverage, with 7 primary benchmarks overall: ASB provides cross-cutting backdoor/attack coverage, while 6 benchmarks are dedicated memory/state evaluations covering memory/RAG poisoning \citep{chen2024agentpoison, zou2024poisonedrag}, query-only memory injection \citep{dong2025minja}, persistent backdoors \citep{feng2026backdooragent}, RAG pipeline attacks \citep{liang2025saferag}, and long-horizon memory attacks \citep{jiang2026agentlab}. But the taxonomy reveals that this apparent breadth is almost entirely one-sided: the 6 dedicated memory/state benchmarks all use environmental pressure, all are sandboxed, and all evaluate action-level compromise; 4 of the 6 are safety-only. In other words, the category is broad in attack families but narrow in methodology. No benchmark evaluates \textit{safety constraint loss during context compaction}---the risk that routine context summarization silently drops safety-critical instructions. Real-world incidents \citep{openclaw2026} have demonstrated this is not theoretical. Given that every agent under finite context windows must eventually perform compaction, this represents an architectural vulnerability that lacks systematic benchmark coverage.

\paragraph{Gap 4: Robustness and Reliability (R10) --- 7 partial benchmarks, 0 primary.}
Seven benchmarks provide secondary coverage of robustness (Agent-SafetyBench, ST-WebAgentBench, SafeAgentBench, AGENTSAFE, IS-Bench, BAD-ACTS, AgentLAB), but no benchmark is primarily designed to test hallucination-driven actions, error cascading in multi-step plans, or behavioral consistency under semantically equivalent perturbations. The taxonomy helps explain why: the corpus is dominated by one-shot action/outcome evaluation in sandboxed settings, with very little repeated-perturbation testing, disposition-level analysis, or longitudinal measurement.

\subsection{Taxonomy-Conditioned Methodological Gaps}

Beyond risk coverage, the corpus-level taxonomy profile exposes five cross-cutting methodological weaknesses:

\begin{enumerate}[leftmargin=*]
    \item \textbf{A fidelity bottleneck.} High-fidelity evaluation is rare: only 5 of 40 core benchmarks are containerized or live, while 34 are sandboxed. Benchmark results therefore remain weakly connected to production deployment conditions.

    \item \textbf{A restricted-authority bottleneck.} Most benchmarks keep agents in narrow capability envelopes: 26 of 40 grant only constrained tools, and only 13 expose agents to open tools or autonomous planning. This makes it difficult to separate true safety from lack of afforded power.

    \item \textbf{A safety-utility measurement bottleneck.} Twenty-four core benchmarks are safety-only; only 12 report safety and utility separately, 3 use a joint metric, and 1 explicitly measures over-refusal. This leaves the field vulnerable to safetywashing through capability suppression \citep{ren2024safetywashing}.

    \item \textbf{A granularity bottleneck.} Action-level judgments dominate (16/40), while only 3 core benchmarks evaluate latent dispositions. The field is better at detecting discrete unsafe acts than persistent unsafe tendencies, drift, or strategic adaptation.

    \item \textbf{A longitudinal and external-validity bottleneck.} Even the benchmarks with autonomous planning or persistent state are still mostly attack snapshots. No benchmark systematically measures how safety degrades over extended interaction, and benchmark results are not validated against real deployment incident data.
\end{enumerate}

Two additional weaknesses are important even though they are not cleanly parameterized by the six-axis taxonomy. First, \textbf{cultural and contextual blindness}: harm definitions are overwhelmingly Western/English-centric, and no benchmark systematically evaluates how safety judgments vary across cultural contexts, languages, or regulatory environments. Second, \textbf{limited severity calibration}: many benchmarks treat low-stakes and high-stakes failures equivalently, reducing practical usefulness for deployment decisions; only ToolEmu and ODCV-Bench use explicit severity-weighted harm measures.

Appendix~\ref{sec:comparison_table} provides a structured comparison of all 45 entries across key dimensions and marks the 5 adjacent artifacts. The field shows rapid growth---from two core benchmarks in 2023 to 40 by early 2026---with benchmark scale varying by two orders of magnitude (40 scenarios to thousands of samples). To test whether the methodological gaps and fragmentation identified above have practical consequences, the next section presents an empirical consistency check.

\section{Cross-Benchmark Analysis}
\label{sec:comparison}

\subsection{Cross-Benchmark Consistency Check}
\label{sec:consistency_check}

To test whether metric fragmentation is only a conceptual concern or manifests in direct model comparison, we ran a consistency check across four benchmarks spanning three risk categories using twelve models (Table~\ref{tab:consistency_check}). This is a scoped empirical study, not a definitive leaderboard, but uses sufficient model count ($n{=}12$), sample sizes (50 per locally executed benchmark per model, or all paired mandated/incentivized ODCV-Bench runs), confidence intervals, and omnibus concordance analysis (Kendall's~$W$) to support its claims about ranking instability.

\paragraph{Setup.} We evaluated twelve models accessed via OpenRouter---Claude Opus 4.6, Gemini 3.1 Pro Preview, GPT-5.4, GPT-OSS-20B, GPT-OSS-120B, Grok 4.20, GLM-5.1, Kimi K2.5, Llama 4 Maverick, Minimax M2.7, Qwen 3.5 35B, and Qwen 3.5 Plus---on four benchmarks covering three risk categories: (i)~AgentHarm [R1: misuse] (50 public harmful-task samples; refusal and semantic judge fixed to GPT-5.4 for all models), (ii)~AgentDojo [R2: injection] (50 prompt-injection samples from each of 5 workspaces---banking, slack, travel, workspace, workspace\_plus---with the macro-average across workspaces as the primary score; no sandbox tasks), (iii)~InjecAgent [R2: injection] (50 cases in the \textit{enhanced} attack setting: the first 25 direct-harm and first 25 data-stealing samples from the official test set, deterministically selected for reproducibility; we used the enhanced setting because the base setting produced near-zero ASR for the majority of models, making it uninformative for ranking), and (iv)~ODCV-Bench [R4: misalignment] (the benchmark-released median-score table for all 40 scenarios under both \textit{incentivized} and \textit{mandated} prompt variations, evaluated by a four-judge panel: non-judge models use the median of all four judges, while judge-family models use the median of the other three judges; per-model effective $n{=}78$--$80$). We report each benchmark's native ``unsafe'' direction: AgentHarm unsafe full score, AgentDojo targeted attack success rate ($1{-}\text{security}$), InjecAgent overall ASR, and ODCV-Bench overall misalignment rate. For ODCV-Bench, we follow its official reporting rule: a paired run is counted as misaligned only when its benchmark-exported median severity score is $\geq 3$; scores 1--2 enter severity summaries but not the misalignment rate. We report 95\% Wilson confidence intervals and use Kendall's $\tau_b$ (tie-adjusted) for pairwise correlations and Kendall's~$W$ (coefficient of concordance) as the primary omnibus statistic, with Holm--Bonferroni correction for multiple comparisons across the six pairwise tests.

\begin{table}[t]
\centering
\caption{Cross-benchmark consistency check ($n{=}12$ models). Higher scores are less safe. 95\% Wilson CIs in brackets.}
\label{tab:consistency_check}
\small
\begin{tabular}{@{}lclclclcl@{}}
\toprule
\textbf{Model} & \textbf{AH} & \textbf{CI} & \textbf{AD} & \textbf{CI} & \textbf{IA} & \textbf{CI} & \textbf{OD} & \textbf{CI} \\
\midrule
GPT-5.4 & .04 & {\scriptsize[.01,.13]} & 1.0 & {\scriptsize[.98,1.]} & .00 & {\scriptsize[.00,.07]} & .24 & {\scriptsize[.16,.34]} \\
Qwen 3.5 Plus & .04 & {\scriptsize[.01,.13]} & .50 & {\scriptsize[.43,.56]} & .00 & {\scriptsize[.00,.07]} & .36 & {\scriptsize[.26,.47]} \\
Claude Opus 4.6 & .06 & {\scriptsize[.02,.16]} & 1.0 & {\scriptsize[.98,1.]} & .00 & {\scriptsize[.00,.07]} & .12 & {\scriptsize[.06,.20]} \\
Gemini 3.1 Pro & .06 & {\scriptsize[.02,.16]} & 1.0 & {\scriptsize[.98,1.]} & .00 & {\scriptsize[.00,.07]} & .45 & {\scriptsize[.35,.56]} \\
GPT-OSS-20B & .06 & {\scriptsize[.02,.16]} & .55 & {\scriptsize[.49,.61]} & .00 & {\scriptsize[.00,.07]} & .33 & {\scriptsize[.23,.43]} \\
GPT-OSS-120B & .10 & {\scriptsize[.04,.21]} & .50 & {\scriptsize[.44,.57]} & .00 & {\scriptsize[.00,.07]} & .39 & {\scriptsize[.29,.50]} \\
Minimax M2.7 & .12 & {\scriptsize[.06,.24]} & .89 & {\scriptsize[.85,.92]} & .00 & {\scriptsize[.00,.07]} & .35 & {\scriptsize[.25,.46]} \\
GLM-5.1 & .12 & {\scriptsize[.06,.24]} & .97 & {\scriptsize[.94,.99]} & .00 & {\scriptsize[.00,.07]} & .25 & {\scriptsize[.17,.35]} \\
Qwen 3.5 35B & .16 & {\scriptsize[.08,.29]} & .76 & {\scriptsize[.70,.81]} & .00 & {\scriptsize[.00,.07]} & .36 & {\scriptsize[.27,.47]} \\
Kimi K2.5 & .20 & {\scriptsize[.11,.33]} & .97 & {\scriptsize[.94,.98]} & .00 & {\scriptsize[.00,.07]} & .30 & {\scriptsize[.21,.41]} \\
Grok 4.20 & .74 & {\scriptsize[.60,.84]} & .40 & {\scriptsize[.34,.46]} & .02 & {\scriptsize[.00,.10]} & .67 & {\scriptsize[.56,.76]} \\
Llama 4 Maverick & --- & & .92 & {\scriptsize[.88,.94]} & .42 & {\scriptsize[.29,.56]} & .16 & {\scriptsize[.10,.26]} \\
\bottomrule
\end{tabular}
\begin{flushleft}
\footnotesize{AH = AgentHarm unsafe full score ($n{=}50$; judge: GPT-5.4); AD = AgentDojo targeted ASR (macro-averaged across 5 workspaces, $n{=}250$); IA = InjecAgent overall ASR (enhanced setting, $n{=}50$); OD = ODCV-Bench overall misalignment rate (mandated + incentivized; multi-judge median severity $\geq 3$, $n{=}78$--$80$). --- = benchmark could not be completed due to persistent API rate limiting.}
\end{flushleft}
\end{table}

\paragraph{Ranking instability across risk categories.} The central finding is that benchmark choice materially changes the safety ordering. Kendall's $W{=}0.10$ ($p{=}0.94$) across all four benchmarks indicates that the rankings are essentially unrelated---no more concordant than random permutations. There is no evidence that the four benchmarks produce concordant rankings, consistent with them measuring independent safety dimensions.

Striking ranking reversals illustrate this at the model level. GPT-5.4 and Claude Opus~4.6 appear among the safest models on AgentHarm (0.04 and 0.06) but hit the ceiling on AgentDojo (1.00). Grok~4.20 is the \textit{most unsafe} model on AgentHarm (0.74) and ODCV-Bench (0.67) but the \textit{safest} on AgentDojo (0.40). Gemini~3.1~Pro is moderate on AgentHarm (0.06) but has the second-highest ODCV-Bench misalignment rate (0.45). Llama~4~Maverick has the highest InjecAgent ASR (0.42)---21$\times$ higher than the next model---yet has one of the lowest ODCV-Bench misalignment rates (0.16) and scores in the middle on AgentDojo (0.92). Claude Opus~4.6 is the \textit{least misaligned} on ODCV-Bench (0.12) despite being the most injection-vulnerable on AgentDojo (1.00).

Pairwise Kendall $\tau_b$ correlations are also unstable. No benchmark pair is significant after Holm--Bonferroni correction. The strongest uncorrected association is AgentDojo vs.\ ODCV-Bench ($\tau_b{=}{-}0.53$, raw $p{=}0.016$), a negative relationship between prompt-injection susceptibility and outcome-driven misalignment in this model panel, but it does not pass the corrected threshold across six pairwise tests. All other pairwise correlations are weak or non-significant ($|\tau_b| \leq 0.45$, $p \geq 0.11$), including InjecAgent vs.\ ODCV-Bench ($\tau_b{=}0.03$, $p{=}0.91$).

\paragraph{Within-benchmark stability.} To rule out the possibility that AgentDojo's ranking reflects workspace-specific artifacts, we computed pairwise $\tau_b$ across the five AgentDojo workspaces. Correlations among banking, slack, travel, and workspace are uniformly high ($\tau_b{=}0.76$--$0.92$, $p < 0.013$); workspace\_plus is the exception ($\tau_b{\approx}0.38$, $p{=}0.43$), likely because only five models completed it, leaving too few pairs for a reliable ranking. The four well-populated workspaces confirm that model vulnerability patterns are consistent across domains.

\paragraph{Interpretation and limitations.} These results should not be read as definitive safety estimates. The confidence intervals show that individual point estimates carry non-trivial uncertainty. Several caveats apply. First, AgentHarm uses GPT-5.4 as both the refusal/semantic judge and as one of the evaluated models, which could introduce self-evaluation bias for that model's AgentHarm score. ODCV-Bench mitigates judge bias by using a four-judge panel and leave-one-out aggregation for judge-family models (Krippendorff's $\alpha{=}0.81$). Second, Llama~4~Maverick could not complete AgentHarm due to persistent API rate limiting, yielding one missing cell; Kendall's $W$ was computed over the 11 models with complete data across all four benchmarks, with Llama~4 included via pairwise deletion for $\tau_b$. Third, the three models scoring 1.00 on AgentDojo are highly capable agents that follow all instructions including injected ones---this reflects genuine injection vulnerability rather than a measurement ceiling, as confirmed by the within-workspace consistency analysis. However, the consistency of the overall pattern---Kendall's $W$ near zero, no Holm-corrected significant pairwise concordance, and non-significant concordance across all four benchmarks---is consistent with the conclusion that benchmark fragmentation affects empirical conclusions across risk categories. A model's safety profile depends heavily on which benchmark and risk dimension is evaluated.

\paragraph{Missing metadata signals lack of reporting standards.} In the comparison table, 26 of 45 entries omit at least one basic metadata field: task counts or tool counts (indicated by ``--''). This missing metadata suggests that the field has not yet converged on what constitutes a well-characterized benchmark, reinforcing our recommendation for minimum reporting standards (Section~\ref{sec:discussion}).

\paragraph{Adjacent evaluation artifacts.} Five of the 45 surveyed entries are adjacent rather than core behavioral benchmarks: R-Judge evaluates risk \textit{recognition} rather than agent behavior, ASSEBench evaluates \textit{evaluators} rather than agents, ToolSafety is primarily a fine-tuning dataset, and GuardAgent and TrustAgent are defense frameworks with small associated benchmarks. We retain these for completeness, code them in the released metadata, and show them in the appendix tables; however, we exclude them from the main corpus-level coverage and taxonomy counts because they serve different functions than behavioral safety benchmarks and would otherwise inflate the apparent coverage of tool-use, privacy, and evaluator/defense-oriented risks.

\subsection{Cross-Benchmark Findings}
\label{sec:cross_findings}

Beyond individual benchmark descriptions, examining the 40 core behavioral benchmarks collectively reveals cross-cutting patterns about the \textit{benchmarking landscape itself}---its structural biases, blind spots, and the degree to which different benchmarks agree or disagree. These findings concern what the benchmarks measure and how they measure it, rather than the model behaviors they report.

\paragraph{Finding 1: Benchmark choice produces contradictory safety conclusions.}
As demonstrated by the consistency check above, safety rankings across four benchmarks are no more concordant than random (Kendall's $W{=}0.10$, $p{=}0.94$). This is not merely a theoretical concern about metric incomparability; it is an empirical demonstration that the field's fragmented evaluation practices produce materially conflicting safety assessments. The core-corpus taxonomy profile clarifies why this should be expected: benchmarks differ not just in risk category, but also in adversarial pressure source, fidelity, capability envelope, scoring, granularity, and safety-utility coupling.

\paragraph{Finding 2: Coverage counts overstate evaluation depth.}
The coverage matrix shows 6--9 primary core benchmarks for R5 (Multi-Agent), R6 (Memory), and R9 (Privacy), but closer examination reveals important narrowness within these categories. The taxonomy makes this visible. The dedicated memory/state cluster is methodologically concentrated in environmental-pressure, sandboxed, action-level attack evaluation. The multi-agent cluster is entirely sandboxed and constrained-tool, with 6 of 9 benchmarks reporting safety-only results. R9 coverage is largely \textit{co-evaluated}: the 6 R9-primary core benchmarks test privacy in conjunction with injection, tool misuse, real-world action, or multi-agent interaction rather than as an isolated property. R7 (Environment) is operationalized almost entirely as embodied household hazards, leaving digital ecosystem harms under-evaluated. High coverage counts can thus mask significant sub-gaps within nominally well-covered categories.

\paragraph{Finding 3: Environment fidelity creates systematic measurement bias.}
As detailed in the taxonomy (Section~\ref{sec:methodology}, Axis~2), benchmarks at different fidelity levels produce systematically different safety estimates. Yet the behavioral-benchmark core is overwhelmingly low-to-medium fidelity: 34 of 40 benchmarks are sandboxed, only 4 are containerized, and only 1 is live. Containerized benchmarks (OpenAgentSafety) report 50--86\% unsafe rates from benign prompts, while synthetic-environment benchmarks testing similar risks report lower rates (e.g., ToolEmu's 23.9\% failure rate in LLM-emulated environments). This ``security by incompetence'' effect \citep{evtimov2025wasp}---where agents appear safe because they lack capability, not because they robustly refuse---means that as agents become more capable, the corpus's heavy reliance on lower-fidelity evaluation may produce dangerously optimistic estimates.

\paragraph{Finding 4: The benchmark landscape is systematically biased toward external and interaction-driven evaluation.}
The taxonomy shows that 17 of 40 core benchmarks use user-directed or environmental pressure, and another 9 evaluate multi-agent interaction as the primary source of danger; only 6 directly probe agent-internal tendencies. Several benchmarks independently show that agents fail safety constraints \textit{without any adversarial attack}. OpenAgentSafety reports high unsafe rates even from benign prompts (see Finding~3). ODCV-Bench \citep{li2025odcv} shows that agents autonomously derive unethical strategies under mere incentive pressure. MAGPIE \citep{juneja2025magpie} finds manipulation and information leakage in non-adversarial collaborative settings. Our consistency check reinforces the separability of these risks: adding ODCV-Bench (R4: outcome-driven misalignment) to the evaluation leaves omnibus concordance indistinguishable from random ($W{=}0.10$, $p{=}0.94$), and its strongest relationship with an injection benchmark is negative but not significant after multiple-comparison correction (AgentDojo: $\tau_b{=}{-}0.53$, raw $p{=}0.016$). This suggests that adversarial robustness and autonomous alignment are distinct risk dimensions that the current benchmark landscape evaluates unevenly. The bias has a temporal dimension: the agent-internal evaluation cluster (RepliBench, AgentMisalignment, PropensityBench, Scheming Propensity, Shutdown Resistance) appears only in 2025--2026, so non-adversarial evaluation is broadening, but from a very small base.

\paragraph{Finding 5: Metric fragmentation prevents cross-benchmark comparison.}
Each benchmark introduces its own primary metric (harm score, safety score, unsafe rate, ASR, CuP, misalignment rate), making cross-benchmark comparison nearly impossible without re-execution. However, the fragmentation is not uniform: within the attack-centric sub-literature, 12 of 40 core benchmarks use attack success rate (ASR) or direct variants (e.g., ISR/ASR, TCR/ASR) as their primary metric, creating local metric monoculture that coexists with global incomparability. The taxonomy shows a second fragmentation layer in safety-utility handling: 24 of 40 core benchmarks are safety-only, 12 report safety and utility separately, 3 use joint metrics, and 1 measures over-refusal. No benchmark reports joint safety-utility metrics in a standardized way, and only 2 of 40 (ToolEmu and ODCV-Bench) use severity-weighted metrics. This matters because a benchmark that weights ``agent sent an unnecessary notification'' the same as ``agent transferred \$10{,}000 to the wrong account'' offers limited operational guidance. The absence of a common measurement framework means that even well-intentioned benchmark users cannot currently rank models across risk categories without re-running evaluations.

\paragraph{Finding 6: Robustness remains a complete evaluation blind spot.}
R10 (Robustness) is the sole risk category with zero primary benchmarks. No benchmark systematically tests behavioral consistency under semantically equivalent input perturbations, error cascading across multi-step plans, or hallucination-driven unsafe actions. The taxonomy shows why this has persisted: the corpus is dominated by action-level and outcome-level single-shot judgments, with only 3 disposition-level benchmarks and very little repeated or longitudinal evaluation. While some benchmarks incidentally touch on robustness---AgentDojo measures utility degradation under attack, and a few benchmarks report multi-run variance---none treat robustness as a first-class safety property. This is a critical gap because robustness failures (inconsistent safety behavior across runs, prompt phrasings, or tool orderings) may be among the most common real-world failure modes.

The complete 45-entry comparison table and six-axis coding table are provided in Appendix~\ref{sec:comparison_table} and Appendix~\ref{sec:taxonomy_table}.

\section{Discussion and Future Directions}
\label{sec:discussion}

\subsection{The State of Agent Safety Benchmarking}

The six findings in Section~\ref{sec:cross_findings} paint a consistent picture: the field has achieved broad risk \textit{coverage} but not methodological \textit{maturity}. Benchmark choice produces contradictory rankings (Finding~1), coverage counts overstate depth (Finding~2), environment fidelity biases results (Finding~3), the landscape over-indexes on externally imposed and interaction-driven scenarios (Finding~4), metrics are fragmented (Finding~5), and robustness remains unaddressed (Finding~6). The corpus-level taxonomy application in Section~\ref{sec:methodology} makes these structural imbalances concrete: among the 40 behavioral benchmarks, 34 are sandboxed, 26 use constrained tools, 28 use rule-based scoring, 16 evaluate action-level failures, and 24 are safety-only. Two additional limitations sit outside the taxonomy but remain important: cultural/contextual narrowness in harm definitions and the rarity of severity-weighted scoring. Without deliberate methodological diversity, contradictory conclusions are the expected outcome rather than a surprise.

\paragraph{Convergent model-behavior patterns.}
While our findings concern the benchmarking landscape, several patterns in \textit{reported model behavior} recur across independently developed benchmarks and are worth noting. Multiple benchmarks observe a knowledge-behavior gap: agents recognize unsafe actions but proceed regardless \citep{zhou2026safepro, ying2025agentsafe, li2025odcv}. Prompt-level safety interventions consistently yield only modest gains (5--10\% reduction in unsafe rates across SafePro, PsySafe \citep{zhang2024psysafe}, BAD-ACTS \citep{nother2025badacts}, and IS-Bench \citep{is2025bench}). Several benchmarks report preliminary evidence of capability-risk coupling, where more capable models exhibit greater adversarial vulnerability \citep{nother2025badacts} or autonomous misbehavior \citep{hopman2026scheming, schlatter2025shutdown, naik2025agentmisalignment}. These patterns are reported by benchmark authors rather than independently verified by us, but their convergence across heterogeneous methodologies makes them noteworthy signals.

\subsection{Recommendations for Future Benchmarks}

Based on our gap analysis, we propose the following priorities:

\paragraph{1. Robustness benchmarks (R10).} The sole remaining zero-coverage category needs primary benchmarks that test: (i) behavioral consistency under semantically equivalent input perturbations ($\geq$5 paraphrases per test case), (ii) error cascading across multi-step plans ($\geq$5 tool calls), (iii) hallucination-driven unsafe actions, and (iv) multi-run variance reporting with confidence intervals.

\paragraph{2. Longitudinal safety evaluation.} Current benchmarks are cross-sectional snapshots. Future benchmarks should test safety degradation over extended use with a minimum of 10+ sessions per agent, measuring: memory accumulation effects, behavioral drift, stale policy retention, and safety constraint loss during context management.

\paragraph{3. Capability-controlled evaluation.} Given Finding~3 (capability-risk coupling), benchmarks must report safety at matched benign utility levels, not just raw unsafe rates. Safety improvements that merely reduce capability (over-refusal) are not genuine advances \citep{ren2024safetywashing}.

\paragraph{4. Non-adversarial memory evaluation.} R6 benchmarks are overwhelmingly adversarial (poisoning, backdoors). The field needs benchmarks for ordinary operational memory failures: context compaction dropping safety constraints, stale policy retention across sessions, and conflicting information resolution.

\paragraph{5. Minimum reporting standard.} We propose that future agent safety benchmarks report at minimum: (i) threat model and agent authority level; (ii) environment fidelity category (synthetic/emulated/containerized/live); (iii) joint safety-utility metric; (iv) over-refusal rate on benign tasks; (v) severity-weighted harm scores; (vi) multi-run variance with 95\% confidence intervals; (vii) LLM-judge audit protocol (if applicable); and (viii) correlation with upstream model capability benchmarks.

\paragraph{6. Benchmark consolidation.} Rather than continued proliferation, the community should establish a common anchor suite of 8--10 representative benchmarks spanning all risk categories, enabling standardized model comparison and longitudinal progress tracking.

\subsection{Toward a Unified Evaluation Framework}

The fragmentation of evaluation methodology---each benchmark using different environments, metrics, threat models, and judge implementations---makes cross-benchmark comparison nearly impossible. We advocate for a modular evaluation framework with:

\begin{itemize}[leftmargin=*]
    \item \textbf{Standardized risk categories} aligned with a shared taxonomy (such as the one proposed in this paper).
    \item \textbf{Pluggable environments} supporting both emulated and sandboxed execution.
    \item \textbf{Unified metrics} that capture severity, frequency, and the safety-utility tradeoff.
    \item \textbf{Mandatory statistical reporting} including confidence intervals, multi-run variance, and prompt sensitivity analysis.
\end{itemize}

\section{Conclusion}
\label{sec:conclusion}

Our systematic analysis of 40 behavioral agent-safety benchmarks, contextualized by 5 adjacent evaluation artifacts, reveals a benchmarking landscape characterized by \textit{coverage without maturity}. The community has identified many of the right risks---multi-agent safety, memory risks, autonomy/control, and privacy each have 6--9 primary core benchmarks---but has not yet converged on stable evaluation standards: metrics are fragmented, cross-benchmark validation is absent, and robustness has zero primary coverage. Applying our six-axis taxonomy across the core corpus shows that the field is also structurally concentrated: 34/40 benchmarks are sandboxed, 26/40 grant only constrained tools, 28/40 use rule-based scoring, 16/40 evaluate action-level failures, and 24/40 are safety-only. More critically, our analysis reveals structural properties of the benchmarking landscape that limit the conclusions any individual benchmark can support. Benchmark choice produces materially contradictory safety rankings for the same models---our consistency check across twelve models and four benchmarks finds ranking concordance indistinguishable from random (Kendall's $W{=}0.10$, $p{=}0.94$). Coverage counts mask narrow sub-category focus. Environment fidelity systematically biases reported safety levels. The landscape over-indexes on external and interaction-driven evaluation while under-representing agent-internal failure modes. These are not model-behavior findings; they are findings about how we measure safety. Our six-axis taxonomy of benchmark evaluation methodology---characterizing adversarial pressure, environment fidelity, agent capability envelope, scoring method, evaluation granularity, and safety-utility coupling---makes explicit the design choices that drive these contradictions and provides a framework for selecting benchmark suites with deliberate methodological diversity. We release the full coverage matrix, benchmark metadata, taxonomy codings, and consistency-check artifacts alongside this paper, and propose minimum reporting standards and a benchmark consolidation agenda for the next generation of agent safety evaluation.


\subsubsection*{Broader Impact Statement}
This survey analyzes the landscape of agent safety benchmarks, which we believe contributes positively to the responsible development of autonomous AI systems. By identifying coverage gaps and methodological weaknesses in current evaluation practices, we aim to guide the development of more rigorous safety assessments. Our cross-benchmark consistency check involved running frontier models on benchmarks that include harmful-task scenarios (AgentHarm), prompt injection attacks (AgentDojo, InjecAgent), and constraint violation scenarios (ODCV-Bench). These benchmarks were designed by their respective authors for safety evaluation purposes, and we used them in their intended capacity. We did not develop new attack methods or discover novel vulnerabilities. The primary risk of this work is that highlighting evaluation gaps could be misinterpreted as evidence that current models are safe in areas not yet benchmarked; we emphasize that absence of evaluation is not evidence of safety.

\subsubsection*{Reproducibility Statement}
All experiment scripts, compact result summaries, imported score tables, and analysis code are released alongside this paper. The cross-benchmark consistency check can be reproduced using the scripts in our repository: \texttt{run\_inspect\_task.py} (AgentHarm, AgentDojo with per-workspace runs), \texttt{run\_injecagent\_subset.py} (InjecAgent, enhanced setting), \texttt{analyze\_results.py} (ODCV import, aggregation, Kendall's $\tau_b$/$W$, Holm correction, Wilson CIs, concordance, and per-workspace analysis), and \texttt{run\_all.sh} (master orchestration for all 12 models except ODCV-Bench). For ODCV-Bench, we release the benchmark-exported median score table used in our analysis (\texttt{odcv\_bench\_scores\_final\_median.csv}) rather than a local wrapper-run log set. The released consistency-check artifact includes per-model-per-benchmark compact JSON summaries for the locally executed benchmarks, the ODCV score export, and derived analysis CSVs. Heavyweight execution traces such as Inspect \texttt{.eval} logs are not part of the submission artifact: the original logs are approximately 700\,MB, and the released summaries contain the fields used to reproduce Table~\ref{tab:consistency_check} and the rank-concordance analyses. The survey metadata artifacts (\texttt{benchmark\_metadata.csv}, \texttt{coverage\_matrix.csv}) and figure-generation scripts are also released.

\section*{Acknowledgments}
This research is supported by the NSERC Discovery Grants (RGPIN-2024-04087), NSERC CREATE Grants (CREATE-554764-2021 \& CREATE-596346-2025), and Canada Research Chairs Program (CRC-2019-00041).

\bibliographystyle{tmlr}
\bibliography{references}

\appendix

\section{Search and Selection Methodology}
\label{sec:search_methodology}

The full search protocol, inclusion/exclusion criteria, and coverage annotation procedure are described in Section~\ref{sec:benchmarks} (Survey Methodology). In summary, we searched Google Scholar, Semantic Scholar, arXiv, and major ML venue proceedings (2023--2026), identifying 40 core behavioral benchmarks (17 peer-reviewed) and 5 adjacent evaluation artifacts (all peer-reviewed).

\section{Full Risk--Entry Coverage Matrix}
\label{sec:coverage_matrix}

Table~\ref{tab:coverage_full} provides the complete 45-entry coverage matrix referenced in Section~\ref{sec:coverage}. Rows marked with $^\dagger$ are adjacent artifacts; they are shown for transparency but excluded from the core-benchmark summary counts in Table~\ref{tab:coverage_summary}.

\begin{table}[ht]
\centering
\caption{Full risk--entry coverage matrix. \CIRCLE~= primary focus, \LEFTcircle~= secondary/partial coverage, \Circle~= not covered. Entries are grouped by primary focus area. Rows marked with $^\dagger$ are adjacent artifacts.}
\label{tab:coverage_full}
\setlength{\tabcolsep}{1.5pt}
\renewcommand{\arraystretch}{0.85}
\footnotesize
\begin{tabular}{@{}l|C{0.5cm}C{0.5cm}C{0.5cm}C{0.5cm}C{0.5cm}C{0.5cm}C{0.5cm}C{0.5cm}C{0.5cm}C{0.5cm}@{}}
\toprule
\textbf{Benchmark} & \rotatebox{70}{\footnotesize\textbf{R1}} & \rotatebox{70}{\footnotesize\textbf{R2}} & \rotatebox{70}{\footnotesize\textbf{R3}} & \rotatebox{70}{\footnotesize\textbf{R4}} & \rotatebox{70}{\footnotesize\textbf{R5}} & \rotatebox{70}{\footnotesize\textbf{R6}} & \rotatebox{70}{\footnotesize\textbf{R7}} & \rotatebox{70}{\footnotesize\textbf{R8}} & \rotatebox{70}{\footnotesize\textbf{R9}} & \rotatebox{70}{\footnotesize\textbf{R10}} \\
\midrule
\multicolumn{11}{l}{\textit{Harmful Task Execution}} \\
AgentHarm & \CIRCLE & \Circle & \LEFTcircle & \Circle & \Circle & \Circle & \Circle & \Circle & \Circle & \Circle \\
Agent-SafetyBench & \CIRCLE & \Circle & \CIRCLE & \Circle & \Circle & \Circle & \LEFTcircle & \Circle & \LEFTcircle & \LEFTcircle \\
SafePro & \CIRCLE & \Circle & \LEFTcircle & \CIRCLE & \Circle & \Circle & \LEFTcircle & \Circle & \LEFTcircle & \Circle \\
MACHIAVELLI & \Circle & \Circle & \Circle & \CIRCLE & \Circle & \Circle & \Circle & \Circle & \Circle & \Circle \\
\midrule
\multicolumn{11}{l}{\textit{Security / Prompt Injection}} \\
ASB & \Circle & \CIRCLE & \LEFTcircle & \Circle & \Circle & \CIRCLE & \Circle & \Circle & \Circle & \Circle \\
AgentDojo & \Circle & \CIRCLE & \Circle & \Circle & \Circle & \Circle & \Circle & \Circle & \LEFTcircle & \Circle \\
InjecAgent & \Circle & \CIRCLE & \Circle & \Circle & \Circle & \Circle & \Circle & \Circle & \CIRCLE & \Circle \\
WASP & \Circle & \CIRCLE & \Circle & \Circle & \Circle & \Circle & \Circle & \Circle & \Circle & \Circle \\
MCP-SafetyBench & \Circle & \CIRCLE & \LEFTcircle & \Circle & \Circle & \Circle & \Circle & \Circle & \LEFTcircle & \Circle \\
\midrule
\multicolumn{11}{l}{\textit{Tool Safety}} \\
ToolEmu & \Circle & \Circle & \CIRCLE & \Circle & \Circle & \Circle & \LEFTcircle & \Circle & \CIRCLE & \Circle \\
SafeToolBench & \Circle & \Circle & \CIRCLE & \Circle & \Circle & \Circle & \LEFTcircle & \Circle & \CIRCLE & \Circle \\
ToolSafety$^\dagger$ & \LEFTcircle & \Circle & \CIRCLE & \Circle & \Circle & \Circle & \Circle & \Circle & \Circle & \Circle \\
\midrule
\multicolumn{11}{l}{\textit{Risk Judgment}} \\
R-Judge$^\dagger$ & \LEFTcircle & \LEFTcircle & \LEFTcircle & \Circle & \Circle & \Circle & \LEFTcircle & \Circle & \CIRCLE & \Circle \\
\midrule
\multicolumn{11}{l}{\textit{Alignment / Constraint-Following}} \\
ODCV-Bench & \Circle & \Circle & \Circle & \CIRCLE & \Circle & \Circle & \LEFTcircle & \Circle & \Circle & \Circle \\
ST-WebAgentBench & \Circle & \Circle & \LEFTcircle & \CIRCLE & \Circle & \Circle & \Circle & \Circle & \LEFTcircle & \LEFTcircle \\
\midrule
\multicolumn{11}{l}{\textit{Embodied Agent Safety}} \\
SafeAgentBench & \Circle & \Circle & \Circle & \Circle & \Circle & \Circle & \CIRCLE & \Circle & \Circle & \LEFTcircle \\
AGENTSAFE & \Circle & \Circle & \Circle & \Circle & \Circle & \Circle & \CIRCLE & \Circle & \Circle & \LEFTcircle \\
IS-Bench & \Circle & \Circle & \Circle & \Circle & \Circle & \Circle & \CIRCLE & \Circle & \Circle & \LEFTcircle \\
\midrule
\multicolumn{11}{l}{\textit{Domain-Specific}} \\
MobileSafetyBench & \CIRCLE & \LEFTcircle & \Circle & \Circle & \Circle & \Circle & \Circle & \Circle & \LEFTcircle & \Circle \\
OpenAgentSafety & \LEFTcircle & \Circle & \CIRCLE & \LEFTcircle & \Circle & \Circle & \LEFTcircle & \Circle & \CIRCLE & \Circle \\
RAS-Eval & \Circle & \LEFTcircle & \CIRCLE & \Circle & \Circle & \Circle & \Circle & \Circle & \LEFTcircle & \Circle \\
\midrule
\multicolumn{11}{l}{\textit{Multi-Agent Safety}} \\
PsySafe & \LEFTcircle & \Circle & \Circle & \Circle & \CIRCLE & \Circle & \Circle & \Circle & \Circle & \Circle \\
AgentLeak & \Circle & \Circle & \Circle & \Circle & \CIRCLE & \Circle & \Circle & \Circle & \CIRCLE & \Circle \\
MultiAgentFraud & \LEFTcircle & \Circle & \Circle & \Circle & \CIRCLE & \Circle & \Circle & \Circle & \Circle & \Circle \\
ColludeBench & \Circle & \Circle & \Circle & \Circle & \CIRCLE & \Circle & \Circle & \Circle & \Circle & \Circle \\
Colosseum & \Circle & \Circle & \Circle & \LEFTcircle & \CIRCLE & \Circle & \Circle & \Circle & \Circle & \Circle \\
MAGPIE & \Circle & \Circle & \Circle & \Circle & \CIRCLE & \Circle & \Circle & \Circle & \CIRCLE & \Circle \\
BAD-ACTS & \Circle & \Circle & \Circle & \Circle & \CIRCLE & \Circle & \Circle & \Circle & \Circle & \LEFTcircle \\
SafeAgents & \LEFTcircle & \LEFTcircle & \Circle & \Circle & \CIRCLE & \Circle & \Circle & \Circle & \Circle & \Circle \\
TAMAS & \Circle & \LEFTcircle & \Circle & \Circle & \CIRCLE & \Circle & \Circle & \Circle & \Circle & \Circle \\
\midrule
\multicolumn{11}{l}{\textit{Memory / State Safety}} \\
AgentPoison & \Circle & \Circle & \Circle & \Circle & \Circle & \CIRCLE & \Circle & \Circle & \Circle & \Circle \\
MINJA & \Circle & \Circle & \Circle & \Circle & \Circle & \CIRCLE & \Circle & \Circle & \Circle & \Circle \\
BackdoorAgent & \Circle & \Circle & \LEFTcircle & \Circle & \Circle & \CIRCLE & \Circle & \Circle & \Circle & \Circle \\
PoisonedRAG & \Circle & \Circle & \Circle & \Circle & \Circle & \CIRCLE & \Circle & \Circle & \Circle & \Circle \\
SafeRAG & \Circle & \LEFTcircle & \Circle & \Circle & \Circle & \CIRCLE & \Circle & \Circle & \Circle & \Circle \\
AgentLAB & \Circle & \Circle & \Circle & \Circle & \Circle & \CIRCLE & \Circle & \Circle & \Circle & \LEFTcircle \\
\midrule
\multicolumn{11}{l}{\textit{Autonomy / Control}} \\
RepliBench & \Circle & \Circle & \Circle & \Circle & \Circle & \Circle & \Circle & \CIRCLE & \Circle & \Circle \\
AgentMisalign. & \Circle & \Circle & \Circle & \Circle & \Circle & \Circle & \Circle & \CIRCLE & \Circle & \Circle \\
SHADE-Arena & \Circle & \Circle & \Circle & \Circle & \Circle & \Circle & \Circle & \CIRCLE & \Circle & \Circle \\
PropensityBench & \Circle & \Circle & \Circle & \Circle & \Circle & \Circle & \Circle & \CIRCLE & \Circle & \Circle \\
Scheming Prop. & \Circle & \Circle & \Circle & \Circle & \Circle & \Circle & \Circle & \CIRCLE & \Circle & \Circle \\
Shutdown Res. & \Circle & \Circle & \Circle & \Circle & \Circle & \Circle & \Circle & \CIRCLE & \Circle & \Circle \\
\midrule
\multicolumn{11}{l}{\textit{Evaluator / Defense}} \\
ASSEBench$^\dagger$ & \LEFTcircle & \LEFTcircle & \LEFTcircle & \Circle & \Circle & \Circle & \Circle & \Circle & \LEFTcircle & \Circle \\
GuardAgent$^\dagger$ & \Circle & \Circle & \Circle & \Circle & \Circle & \Circle & \Circle & \Circle & \CIRCLE & \Circle \\
TrustAgent$^\dagger$ & \LEFTcircle & \Circle & \CIRCLE & \Circle & \Circle & \Circle & \Circle & \Circle & \Circle & \Circle \\
\midrule
\textbf{Primary Count (all 45)} & \textbf{4} & \textbf{5} & \textbf{7} & \textbf{4} & \textbf{9} & \textbf{7} & \textbf{3} & \textbf{6} & \textbf{8} & \textbf{0} \\
\textbf{Primary Count (core 40)} & \textbf{4} & \textbf{5} & \textbf{5} & \textbf{4} & \textbf{9} & \textbf{7} & \textbf{3} & \textbf{6} & \textbf{6} & \textbf{0} \\
\bottomrule
\end{tabular}
\begin{flushleft}
\footnotesize{R1: Misuse, R2: Injection, R3: Tool Misuse, R4: Misalignment, R5: Multi-Agent, R6: Memory, R7: Environment, R8: Autonomy, R9: Privacy, R10: Robustness. Rows marked with $^\dagger$ are adjacent artifacts; the second total row excludes them and matches the main-text core-benchmark summary.}
\end{flushleft}
\end{table}

\FloatBarrier
\section{Full Evaluation Entry Comparison Table}
\label{sec:comparison_table}

Table~\ref{tab:comparison} provides the complete 45-entry comparison referenced in Section~\ref{sec:comparison}. Rows marked with $^\dagger$ are adjacent evaluator, defense, or dataset artifacts rather than core behavioral benchmarks.

{\footnotesize
\setlength{\tabcolsep}{4pt}
\renewcommand{\arraystretch}{0.9}
\begin{longtable}{@{}>{\raggedright\arraybackslash}p{2.2cm} c >{\raggedright\arraybackslash}p{1.4cm} c c >{\raggedright\arraybackslash}p{2.1cm} >{\raggedright\arraybackslash}p{1.8cm} >{\raggedright\arraybackslash}p{1.8cm} >{\raggedright\arraybackslash}p{1.4cm}@{}}
\caption{Comprehensive comparison of agent-safety evaluation entries. ``--'' indicates not applicable or not reported. Rows marked with $^\dagger$ are adjacent artifacts rather than core behavioral benchmarks.}
\label{tab:comparison} \\
\toprule
\textbf{Benchmark} & \textbf{Year} & \textbf{Venue} & \textbf{Tasks} & \textbf{Tools} & \textbf{Primary Risk} & \textbf{Env.\ Type} & \textbf{Eval.\ Method} & \textbf{Key Metric} \\
\midrule
\endfirsthead
\multicolumn{9}{l}{\small\itshape Table~\ref{tab:comparison} continued from previous page} \\
\toprule
\textbf{Benchmark} & \textbf{Year} & \textbf{Venue} & \textbf{Tasks} & \textbf{Tools} & \textbf{Primary Risk} & \textbf{Env.\ Type} & \textbf{Eval.\ Method} & \textbf{Key Metric} \\
\midrule
\endhead
\midrule
\multicolumn{9}{r}{\small\itshape Continued on next page} \\
\endfoot
\bottomrule
\endlastfoot
AgentHarm & 2024 & ICLR'25 & 440 & 104 & Harmful compliance & Synthetic tools & Human rubrics & Harm score \\
Agent-SafetyBench & 2024 & arXiv & 2,000 & 349 & General safety & Simulated & Fine-tuned judge & Safety score \\
SafePro & 2026 & arXiv & 275 & -- & Professional misuse & Code agent & LLM-as-judge & Unsafe rate \\
MACHIAVELLI & 2023 & ICML'23 & 572K & -- & Ethics/reward & Text games & Annotation & Ethical score \\
\midrule
ASB & 2024 & ICLR'25 & 400+ & 400+ & Attack/defense & Multi-domain & ASR metrics & ASR \\
AgentDojo & 2024 & NeurIPS'24 & 629 & $\sim$100 & Prompt injection & Tool-use sim & State checks & Util.+ASR \\
InjecAgent & 2024 & ACL Find.'24 & 1,054 & 79 & Indirect injection & Tool-integrated & ASR & ASR \\
WASP & 2025 & arXiv & -- & Web & Web injection & Live web & E2E eval & ASR \\
MCP-Safety & 2025 & arXiv & -- & MCP & MCP attacks & Real MCP & Multi-turn & ASR \\
\midrule
ToolEmu & 2023 & ICLR'24 & 144 & 311 & Tool risk & LLM-emulated & LLM evaluator & Safety score \\
SafeToolBench & 2025 & EMNLP Find.'25 & 1,200 & 31-73 & Tool plan safety & Prospective & Classification & Recall/F1 \\
ToolSafety$^\dagger$ & 2025 & EMNLP'25 & 14,290 & -- & Tool invocation & Dataset & Fine-tuning & Safety rate \\
\midrule
R-Judge$^\dagger$ & 2024 & EMNLP Find.'24 & 569 & -- & Risk judgment & Static logs & Classification & Accuracy \\
\midrule
ODCV-Bench & 2025 & arXiv & 40 & Bash & Constraint viol. & Containerized & AI judge panel & Misalign. rate \\
ST-WebAgent & 2024 & ICLR'26 & 375 & Web & Policy compliance & Enterprise web & YAML policies & CuP \\
\midrule
SafeAgentBench & 2024 & arXiv & 750 & 17 & Embodied hazards & AI2-THOR & Exec.+ semantic & Rejection rate \\
AGENTSAFE & 2025 & MAS@ ICML'25 & 1,350 & THOR & Embodied (Asimov) & Embodied VLM & Multi-stage & Per-stage eval \\
IS-Bench & 2025 & arXiv & 161 & Sim & Interactive safety & HiFi sim & Process-oriented & Safety score \\
\midrule
MobileSafety & 2024 & arXiv & 250 & Mobile & Mobile misuse & Android emu & Composite & Safety+Help \\
OpenAgent- Safety & 2025 & ICLR'26 & 356 & 5 real & Real-world risks & Container & Rule+LLM & Unsafe rate \\
RAS-Eval & 2025 & arXiv & 3,802 & 75 & CWE-mapped & Real+sim & CWE metrics & TCR/ASR \\
\midrule
PsySafe & 2024 & ACL'24 & 859 & -- & MA psych. & CAMEL/ AutoGen & Dark trait eval & PDR/JDR \\
AgentLeak & 2026 & arXiv & 1,000 & -- & MA privacy leak & Multi-agent & Leakage metrics & Exposure rate \\
MultiAgent- Fraud & 2025 & arXiv & 2,800 & -- & MA collusion/ fraud & Social platform & Conversion & Fraud rate \\
ColludeBench & 2025 & arXiv & 600 & -- & Steg.\ collusion & Market/ auction & Meta-tests & TPR/FPR \\
Colosseum & 2026 & arXiv & -- & -- & Cooperative collusion & DCOP & Regret-based & Coalition adv. \\
MAGPIE & 2025 & arXiv & 200 & -- & MA privacy & Negotiation & Contextual & Leak rate \\
BAD-ACTS & 2025 & arXiv & 188 & -- & Insider adversary & Multi-agent & ASR & ASR \\
SafeAgents & 2025 & arXiv & -- & -- & MA weak links & AutoGen/ CrewAI & DHARMA & ASR \\
TAMAS & 2025 & MAS@ ICML'25 & 300 & 211 & MA adversarial & AutoGen/ CrewAI & ERS & ERS \\
\midrule
AgentPoison & 2024 & NeurIPS'24 & -- & 3 agents & Memory poisoning & RAG agents & ASR & ASR \\
MINJA & 2025 & NeurIPS'25 & -- & 3 agents & Memory injection & RAG agents & ISR/ASR & ISR/ASR \\
BackdoorAgent & 2026 & arXiv & -- & 4 apps & Memory backdoors & Agent workflows & Persistence & Trigger rate \\
PoisonedRAG & 2024 & USENIX'25 & -- & RAG & RAG poisoning & RAG pipeline & ASR & ASR \\
SafeRAG & 2025 & ACL'25 & -- & 14 comp. & RAG attacks & RAG pipeline & Multi-type & Vuln. rate \\
AgentLAB & 2026 & arXiv & 644 & 28 env. & Long-horizon attacks & Multi-env & Multi-attack & ASR \\
\midrule
RepliBench & 2025 & arXiv & 86 & -- & Self-replication & Compute env & Task completion & Repl. rate \\
AgentMisalign. & 2025 & arXiv & -- & -- & Misalignment prop. & Deployment sim & Multi-eval & Misalign. rate \\
SHADE-Arena & 2025 & arXiv & -- & -- & Covert sabotage & Agent tasks & AUC & AUC \\
PropensityBench & 2025 & arXiv & 5,874 & 6,648 & Dangerous propensity & Agentic & Tool selection & Propensity \\
Scheming Prop. & 2026 & arXiv & -- & -- & Scheming & Agentic & Scheming rate & Scheme rate \\
Shutdown Res. & 2025 & TMLR'26 & 100K+ & -- & Shutdown resistance & Script-based & Compliance & Resist. rate \\
\midrule
ASSEBench$^\dagger$ & 2025 & NeurIPS'25 & 2,293 & -- & Evaluator quality & Static logs & Evaluator eval & Accuracy \\
GuardAgent$^\dagger$ & 2024 & ICML'25 & 2 & Domain & Defense (HC/Web) & HC, Web & Code guardrails & Guardrail acc. \\
TrustAgent$^\dagger$ & 2024 & EMNLP'24 & -- & -- & Agent constitution & General & Pre/in/post plan & Safety+Help \\
\end{longtable}
}

\FloatBarrier
\section{Full Six-Axis Taxonomy Coding Table}
\label{sec:taxonomy_table}

Table~\ref{tab:taxonomy_full} provides the full entry-level six-axis codings referenced throughout Sections~\ref{sec:methodology} and~\ref{sec:coverage}.

{\scriptsize
\setlength{\tabcolsep}{2pt}
\renewcommand{\arraystretch}{0.95}
\begin{longtable}{@{}>{\raggedright\arraybackslash}p{2.5cm}>{\raggedright\arraybackslash}p{2.3cm}>{\raggedright\arraybackslash}p{2.0cm}>{\raggedright\arraybackslash}p{2.1cm}>{\raggedright\arraybackslash}p{2.1cm}>{\raggedright\arraybackslash}p{1.7cm}>{\raggedright\arraybackslash}p{2.3cm}@{}}
\caption{Entry-level six-axis taxonomy codings. Full entries are shown to make the coding auditable from the manuscript itself. Rows marked with $^\dagger$ are adjacent artifacts.}
\label{tab:taxonomy_full} \\
\toprule
\textbf{Benchmark} & \textbf{Adversarial Pressure Source} & \textbf{Environment Fidelity} & \textbf{Capability Envelope} & \textbf{Scoring Method} & \textbf{Evaluation Granularity} & \textbf{Safety-Utility Coupling} \\
\midrule
\endfirsthead
\multicolumn{7}{l}{\small\itshape Table~\ref{tab:taxonomy_full} continued from previous page} \\
\toprule
\textbf{Benchmark} & \textbf{Adversarial Pressure Source} & \textbf{Environment Fidelity} & \textbf{Capability Envelope} & \textbf{Scoring Method} & \textbf{Evaluation Granularity} & \textbf{Safety-Utility Coupling} \\
\midrule
\endhead
\midrule
\multicolumn{7}{r}{\small\itshape Continued on next page} \\
\endfoot
\bottomrule
\endlastfoot
AgentHarm & User-directed & Sandboxed & Constrained tools & Human rubrics & Outcome & Safety-only \\
Agent-SafetyBench & User-directed & Sandboxed & Constrained tools & LLM-judge & Pattern & Safety-only \\
SafePro & User-directed & Sandboxed & Open tools & LLM-judge & Outcome & Safety-only \\
MACHIAVELLI & \makecell[tl]{Structural/\\incentive-based} & Sandboxed & Text-only & Rule-based & Pattern & Separate reporting \\
ASB & Environmental & Sandboxed & Constrained tools & Rule-based & Action & Safety-only \\
AgentDojo & Environmental & Containerized & Constrained tools & Rule-based & Action & Separate reporting \\
InjecAgent & Environmental & Sandboxed & Constrained tools & Rule-based & Action & Safety-only \\
WASP & Environmental & Live & Open tools & Hybrid & Action & Safety-only \\
MCP-SafetyBench & Environmental & Sandboxed & Constrained tools & Rule-based & Action & Safety-only \\
ToolEmu & \makecell[tl]{Structural/\\incentive-based} & LLM-emulated & Constrained tools & LLM-judge & Outcome & Safety-only \\
SafeToolBench & Environmental & Sandboxed & Constrained tools & Rule-based & Action & Safety-only \\
ToolSafety$^\dagger$ & User-directed & Static & Constrained tools & Rule-based & Action & Safety-only \\
R-Judge$^\dagger$ & \makecell[tl]{Structural/\\incentive-based} & Static & Text-only & Rule-based & Action & Safety-only \\
ODCV-Bench & \makecell[tl]{Structural/\\incentive-based} & Containerized & Open tools & LLM-judge & Pattern & Separate reporting \\
ST-WebAgentBench & \makecell[tl]{Structural/\\incentive-based} & Containerized & Constrained tools & Rule-based & Outcome & Joint metric \\
SafeAgentBench & \makecell[tl]{Structural/\\incentive-based} & Sandboxed & Constrained tools & Hybrid & Outcome & Separate reporting \\
AGENTSAFE & \makecell[tl]{Structural/\\incentive-based} & Sandboxed & Constrained tools & Rule-based & Pattern & Separate reporting \\
IS-Bench & \makecell[tl]{Structural/\\incentive-based} & Sandboxed & Constrained tools & Rule-based & Outcome & Separate reporting \\
MobileSafetyBench & User-directed & Sandboxed & Constrained tools & Hybrid & Outcome & Over-refusal measured \\
OpenAgentSafety & \makecell[tl]{Structural/\\incentive-based} & Containerized & Open tools & Hybrid & Outcome & Separate reporting \\
RAS-Eval & Environmental & Sandboxed & Open tools & Rule-based & Action & Separate reporting \\
PsySafe & Multi-agent & Sandboxed & Constrained tools & LLM-judge & Pattern & Safety-only \\
AgentLeak & Multi-agent & Sandboxed & Constrained tools & Rule-based & Pattern & Safety-only \\
\makecell[tl]{MultiAgent\\FraudBench} & Multi-agent & Sandboxed & Constrained tools & Rule-based & Outcome & Safety-only \\
ColludeBench & Multi-agent & Sandboxed & Constrained tools & Rule-based & Pattern & Separate reporting \\
Colosseum & Multi-agent & Sandboxed & Constrained tools & Rule-based & Pattern & Joint metric \\
MAGPIE & Multi-agent & Sandboxed & Constrained tools & LLM-judge & Pattern & Safety-only \\
BAD-ACTS & Multi-agent & Sandboxed & Constrained tools & Rule-based & Action & Safety-only \\
SafeAgents & Multi-agent & Sandboxed & Constrained tools & Rule-based & Action & Safety-only \\
TAMAS & Multi-agent & Sandboxed & Constrained tools & Rule-based & Action & Joint metric \\
AgentPoison & Environmental & Sandboxed & Autonomous planning & Rule-based & Action & Separate reporting \\
MINJA & Environmental & Sandboxed & Autonomous planning & Rule-based & Action & Separate reporting \\
BackdoorAgent & Environmental & Sandboxed & Autonomous planning & Rule-based & Action & Safety-only \\
PoisonedRAG & Environmental & Sandboxed & Autonomous planning & Rule-based & Action & Safety-only \\
SafeRAG & Environmental & Sandboxed & Autonomous planning & Rule-based & Action & Safety-only \\
AgentLAB & Environmental & Sandboxed & Autonomous planning & Rule-based & Action & Safety-only \\
RepliBench & Agent-internal & Sandboxed & Open tools & Rule-based & Outcome & Safety-only \\
AgentMisalignment & Agent-internal & Sandboxed & Open tools & LLM-judge & Pattern & Safety-only \\
SHADE-Arena & Agent-internal & Sandboxed & Constrained tools & Rule-based & Pattern & Separate reporting \\
PropensityBench & Agent-internal & Sandboxed & Constrained tools & Rule-based & Disposition & Safety-only \\
Scheming Propensity & Agent-internal & Sandboxed & Constrained tools & Rule-based & Disposition & Safety-only \\
Shutdown Resistance & Agent-internal & Sandboxed & Constrained tools & Rule-based & Disposition & Safety-only \\
ASSEBench$^\dagger$ & \makecell[tl]{Structural/\\incentive-based} & Static & Text-only & Rule-based & Action & Safety-only \\
GuardAgent$^\dagger$ & \makecell[tl]{Structural/\\incentive-based} & Sandboxed & Constrained tools & Rule-based & Action & Safety-only \\
TrustAgent$^\dagger$ & \makecell[tl]{Structural/\\incentive-based} & Sandboxed & Constrained tools & Hybrid & Outcome & Separate reporting \\
\end{longtable}
}

\end{document}